\documentclass{IEEEtaes}
\usepackage{longtable}
\usepackage{todonotes}
\usepackage{cuted}
\graphicspath{ {./images/} }

\usepackage[figuresright]{rotating}
\usepackage{xy}
\usepackage{booktabs}
\usepackage{pifont}
\usepackage{color}
\usepackage{listings}
\usepackage{pdfpages}
\usepackage{array}
\usepackage{algorithm}
\usepackage{algpseudocode}
\usepackage{float}
\usepackage{caption}
\usepackage{lastpage}
\usepackage{afterpage}
\usepackage{lipsum}
\usepackage{adjustbox}
\usepackage{rotating}

\algnewcommand\algorithmicinput{\textbf{Output:}}
\algnewcommand\Output{\item[\algorithmicinput]}

\let\oldReturn\Return
\renewcommand{\Return}{\State\oldReturn}

\usepackage{pifont}
\usepackage{amsmath,amssymb}
\usepackage{color}
\usepackage{cite}

\DeclareGraphicsExtensions{.pdf,.png,.jpg}

\usepackage{amsmath}
\usepackage[makeroom]{cancel}
\usepackage{amsthm}
\usepackage{siunitx}
\usepackage{textcomp}
\usepackage{subcaption}

\DeclareSymbolFont{boldoperators}{OT1}{cmr}{bx}{n}
\SetSymbolFont{boldoperators}{bold}{OT1}{cmr}{bx}{n}

\usepackage{tikz}
\usepackage{mathtools}

\usepackage[long]{optidef}

\def\BibTeX{{\rm B\kern-.05em{\sc i\kern-.025em b}\kern-.08em
    T\kern-.1667em\lower.7ex\hbox{E}\kern-.125emX}}
\usepackage{balance}

\usepackage{ragged2e}
\algnewcommand{\LeftComment}[1]{\Statex \(\triangleright\) #1}

\usepackage{setspace}

\usepackage[hang,flushmargin,multiple]{footmisc}

\usepackage{makecell}
\usepackage{multirow}

\usepackage{rotating}
\usepackage{tabularray}

\usepackage{accents}
\newlength{\dhatheight}

\makeatletter
\newcommand\fs@betterruled{%
  \def\@fs@cfont{\bfseries}\let\@fs@capt\floatc@ruled
  \def\@fs@pre{\vspace*{5pt}\hrule height.8pt depth0pt \kern2pt}%
  \def\@fs@post{\kern2pt\hrule\relax}%
  \def\@fs@mid{\kern2pt\hrule\kern2pt}%
  \let\@fs@iftopcapt\iftrue}
\floatstyle{betterruled}
\restylefloat{algorithm}
\makeatother
\usepackage[dvipsnames]{xcolor}
\newcommand{\LRT}[2]{%
  \mathrel{\mathop\gtrless\limits^{#1}_{#2}}%
}

\usepackage{stfloats} 

\usepackage{xcolor}

\usepackage[capitalise]{cleveref}

\begin{document}

\title{An Efficient Blind Detection of OFDM Signals with Unknown Parameters}

\author{Ata Bartu Soyuer}
\affil{ASELSAN Inc., Ankara, Türkiye}

\author{Gokhan M.~Guvensen}
\member{Member, IEEE}
\affil{Middle East Technical University (METU), Ankara, Türkiye}

\receiveddate{This work has been submitted to the IEEE for
    possible publication. Copyright may be transferred
    without notice, after which this version may no
    longer be accessible.
This work was supported by ASELSAN Communication and Information Technologies Sector under the Military Communications System Engineering Directorate.}


\authoraddress{Authors’ address: Ata Bartu Soyuer is with ASELSAN Inc. Communication and Information Technologies Sector, Ankara 06200, Türkiye. (e-mail: \href{mailto:abartusoyuer@aselsan.com}{abartusoyuer@aselsan.com} Gokhan M.~Guvensen is with Electrical and Electronics Engineering Department, Middle East Technical University (METU), Ankara 06800, Türkiye, (e-mail: \href{mailto:guvensen@metu.edu.tr}{guvensen@metu.edu.tr}). {\itshape (Corresponding author: Gokhan M.~Guvensen)}}

\markboth{SOYUER AND GUVENSEN}{BLIND DETECTION OF OFDM SIGNALS}
\maketitle

\begin{abstract}
This study investigates efficient detector designs for the blind detection of orthogonal frequency-division multiplexing (OFDM) signals with unknown power, bandwidth, and carrier frequency in wideband spectrum sensing. We first describe the received signal model and conduct an asymptotic covariance analysis to obtain a tractable discrete-time representation. Based on this model, generalized likelihood ratio test (GLRT) detectors are derived for both known and unknown noise power cases in a single-antenna receiver setting.
The proposed framework is subsequently extended to multi-antenna receivers, and the corresponding multi-antenna GLRT detectors are developed for both known and unknown noise power cases. Importantly, using the cumulative-sum (CUSUM) operations, the resulting detectors reduce to an efficient one-dimensional (1D) search in both single-antenna and multi-antenna settings. We also provide theoretical analysis for the conventional energy detector (Conv-ED) and the hypothetical uniformly most powerful (UMP) test to reveal the impact of system parameters on detection performance. In addition, computational complexity analysis is presented for both the proposed and conventional detectors. Finally, comprehensive simulation results are provided to evaluate the probability of detection. The results demonstrate that the proposed detectors achieve performance close to that of a genie-aided detector while significantly outperforming conventional detection approaches.
\end{abstract}


\section{INTRODUCTION}
Spectrum sensing refers to the detection of signal activity within a frequency band and plays an important role in systems that require dynamic spectrum awareness, such as cognitive radio and adaptive radar. The majority of existing work addresses spectrum sensing in the context of cognitive radio, where it enables the the identification of spectrum opportunities for secondary access \cite{Survey_1,Survey_2,Survey_3}.
%

Spectrum sensing methods can be categorized into knowledge-aided and blind approaches \cite{Survey_1,Survey_2,Survey_3}. Knowledge-aided methods, assume prior knowledge of the signal structure and are therefore unsuitable for scenarios involving unknown signal characteristics.
Spectrum sensing methods are also categorized as wideband (WB) or narrowband (NB), depending on the bandwidth of the frequency region under observation \cite{Survey_1,Survey_2,Survey_3}. WB sensing aims to identify spectrum occupancy over a wide frequency range and typically entails higher hardware and computational complexity, whereas NB sensing focuses on monitoring a single subband. Within the NB framework, the literature considers energy-based, cyclostationary, and covariance-based detectors \cite{Survey_1}. 
%

Energy detection is among the most widely studied spectrum sensing techniques due to its low complexity \cite{ED_1,ED_5,ED_14,TAES_ED_1,TAES_ED_2}. 
%
%
However, its performance degrades in WB sensing scenarios because the energy calculation is dominated by noise over the largely unoccupied spectrum.

Many man-made signals exhibit inherent periodicities related to parameters such as the symbol rate. Cyclostationary-based detectors exploit these properties for signal detection \cite{cyclo_2,cyclo_3}.
%
%
These methods typically rely on the cyclic autocorrelation function (CAF) or its Fourier-domain representation, the spectral correlation function (SCF), which entails relatively high computational complexity due to their 2D structure. 
%

Covariance-based spectrum sensing methods have also been widely investigated \cite{cov_2,cov_3,cov_4,cov_5,cov_6}. These approaches exploit the statistical structure of the received signal, typically through eigenvalue-based tests applied to the sample covariance matrix. Although they offer improved robustness compared to energy detection, eigenvalue-based methods generally incur higher complexity due to the need for eigenvalue decomposition.

Wideband spectrum sensing involves monitoring a broad frequency range and is commonly categorized into Nyquist-based and sub-Nyquist-based approaches \cite{Survey_1,Survey_2}. Nyquist-rate sensing requires higher sampling rates and, consequently, leads to increased requirements on the analog-to-digital converter (ADC). In contrast, sub-Nyquist methods aim to alleviate this burden by reducing the sampling rate while preserving sensing capability.

Sub-Nyquist wideband spectrum sensing aims to detect spectral activity from samples acquired below the Nyquist rate. Most existing approaches rely on compressed sensing, exploiting spectrum sparsity to enable reliable sensing from partial observations \cite{subNyquist_1,subNyquist_3,subNyquist_4}. Although these methods relax ADC and sampling requirements, they incur additional computational complexity due to the sparse recovery stage.

Nyquist-based wideband spectrum sensing acquires signals at or above the Nyquist rate and is primarily constrained by the cost and power consumption of high-speed ADCs. Despite these limitations, such approaches are often preferred in scenarios requiring reliable detection over short observation intervals. The literature commonly considers fast Fourier transform (FFT)-based and wavelet-based methods for Nyquist-rate wideband sensing \cite{WB_1,WB_3,WB_4,wavelet_5,TAES_FFT}. FFT-based techniques benefit from efficient implementations and support various detection strategies, including peak detection, power spectral density (PSD)–based tests, and likelihood-based formulations.

Another class of Nyquist-based approaches relies on channelized radiometers, which employ parallel filter-bank structures to decompose the wideband signal into multiple narrowband subbands \cite{filter_bank_1,filter_bank_2}. This architecture allows each subband to be sampled at a lower rate but requires multiple ADCs operating in parallel. Detection can then be performed independently on each subband using narrowband sensing techniques.
%

Sweeping spectrum analyzers constitute another class of Nyquist-based wideband sensing methods, in which a tunable narrowband filter sequentially scans the frequency range of interest \cite{sweep_1,sweep_2}. By measuring the power in each subband over time, spectral occupancy can be inferred while reducing the instantaneous bandwidth and ADC rate requirements. However, because the spectrum is not observed simultaneously, sweeping-based approaches incur increased latency, which limits their practicality.

Wavelet transform–based spectrum sensing exploits the joint time–frequency localization of wavelets to analyze wideband signals across multiple scales \cite{wavelet_5,wavelet_7}. Owing to their ability to capture sharp spectral transitions, these methods are effective at identifying boundaries between occupied and unoccupied frequency bands. However, compared to FFT, wavelet-based methods generally incur higher computational complexity.

Recent advances in machine learning (ML) and deep learning (DL) have motivated a growing body of work on ML/DL-enhanced spectrum sensing \cite{ML_1,ML_2,ML_3,ML_4}. Rather than forming a distinct class of spectrum sensing methods, these approaches typically build upon classical sensing concepts, such as cyclostationary features, spectrograms, or covariance statistics, and integrate them with ML or DL models to enhance detection performance. Despite the reported performance gains, these methods generally incur increased computational complexity.

\begin{figure}[ht]
    \vspace{1mm}
    \centering
    \includegraphics[width=\columnwidth]{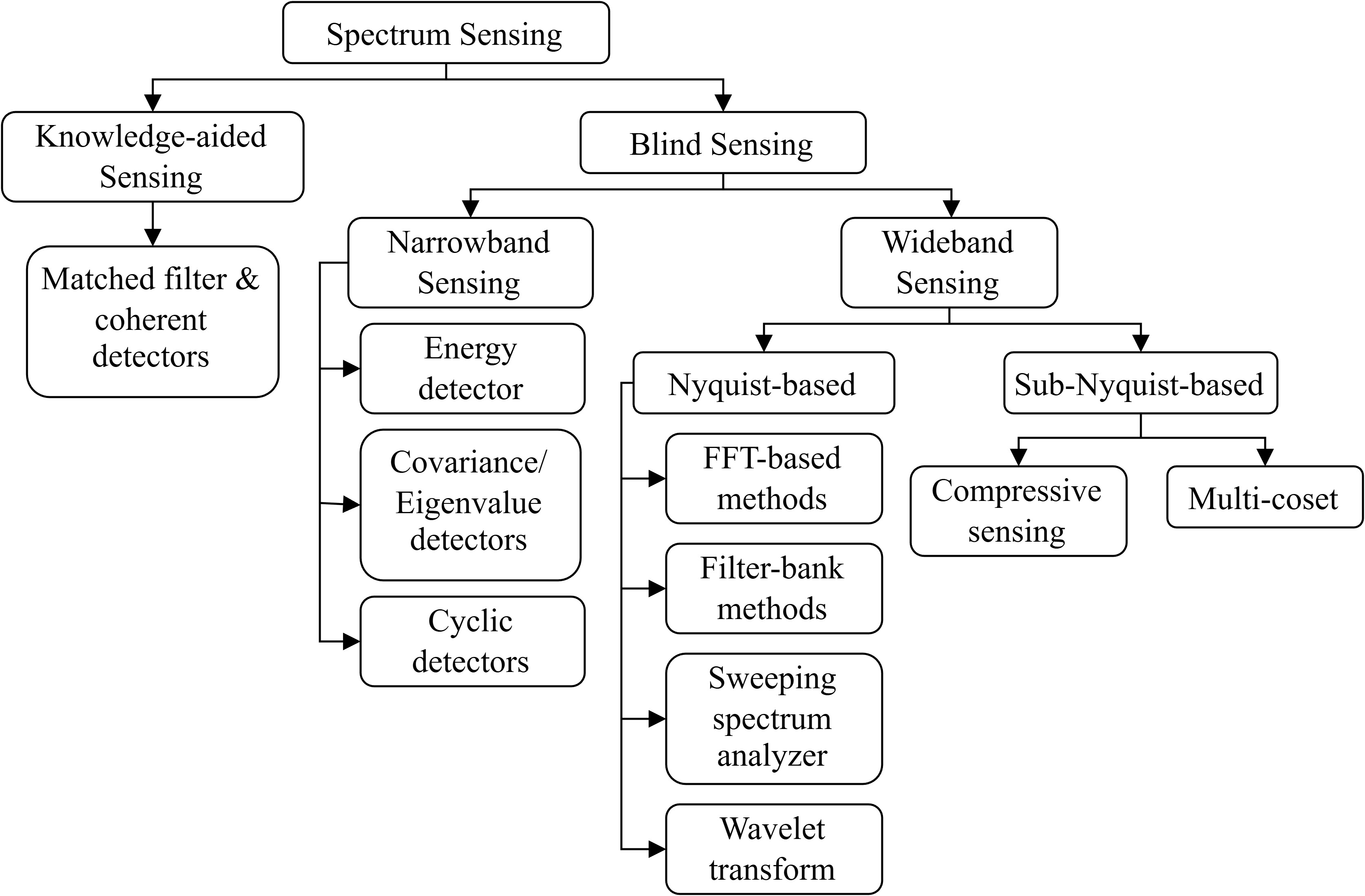}
    \caption{Taxonomy of spectrum sensing methods in the literature}
    \label{fig_scan}
\end{figure}

When monitoring a large band, the receiver collects a significant amount of ambient noise over the entire band. Consequently, in wideband sensing scenarios, the total noise power can easily overwhelm the signal power, leading to a very low signal-to-noise ratio (SNR). To address this challenge, this study proposes GLRT detectors for the blind detection of OFDM signals with unknown parameters in wideband spectrum sensing. We consider both known and unknown noise power scenarios for single-antenna and multi-antenna configurations. To ensure computational efficiency, we develop an algorithm based on an effective 1D search using cumulative sum (CUSUM) operations. While CUSUM is traditionally employed for change-point detection in time series \cite{CUSUM_1,CUSUM_2}, here it is used as a tool to enable the efficient implementation of the proposed detectors.

Among existing works, \cite{GLRT_paper} proposes a related GLRT-based approach. However, their detector compares each candidate bandwidth against a separate threshold, whereas a GLRT, by formulation, is expected to employ a single detection threshold. Moreover, \cite{GLRT_paper} defines SNR as the ratio of signal power density to noise power density. Under this definition, detection performance improves with increasing signal bandwidth; however, this behavior arises from their SNR definition rather than from the detector itself.
In contrast, we define SNR as the ratio of total signal power to total noise power. Under this definition, our results show that detection performance degrades as the bandwidth increases, which is consistent with the principle of direct-sequence spread-spectrum (DSSS) systems. Additionally, our framework extends to multi-antenna receivers, a case not considered in \cite{GLRT_paper}.
The key contributions of this study are summarized as follows:
\begin{itemize} 
    \item We present a covariance analysis showing that the discrete Fourier transform (DFT) samples of the received signal are asymptotically uncorrelated. This analysis enables the formulation of a tractable discrete-time signal representation for blind sensing.

    \item We propose GLRT detectors for both known and unknown noise power cases in the single-antenna receiver scenario. We then extend the proposed detectors to a multi-antenna receiver scenario. Using CUSUM, the proposed GLRT detectors reduce to an effective 1D search over the interval length.

    \item We provide a theoretical analysis of the conventional energy detector (Conv-ED) and the hypothetical uniformly most powerful (UMP) test. This analysis reveals how system parameters influence detection performance and highlights the limitations of the Conv-ED in wideband sensing.
    
    \item We provide comprehensive simulation results evaluating detection performance in terms of the false alarm rate, SNR, bandwidth of the target signal, sampling rate, and number of antennas. The results show that the proposed detectors achieve performance close to that of a genie-aided detector while significantly outperforming conventional approaches.
\end{itemize}

The remainder of this paper is organized as follows. Section~\ref{sec_system_model} presents the system model. Section~\ref{sec_detector_designs} introduces the proposed GLRT detector designs for the single-antenna receiver. Section~\ref{sec_multiantenna_extension} extends the proposed GLRT framework to multi-antenna receivers. Section~\ref{sec_theoretical_analysis} provides theoretical analysis of detection performance. Section~\ref{sec_complexity} analyzes the computational complexity. Finally, Section~\ref{sec_sim_results} presents the simulation results.

\vspace{-3mm}
\subsection{Notations}
Fonts $x$, $\textbf{x}$, and $\textbf{X}$ denote scalars, column vectors, and matrices, respectively. The complex conjugate and magnitude of $x$ are denoted by $x^*$ and $|x|$, respectively. For a matrix $\textbf{X}$, the determinant, inverse, transpose, and Hermitian are denoted by $\operatorname{det}|\textbf{X}|$, $\textbf{X}^{-1}$, $\textbf{X}^T$, and $\textbf{X}^H$, respectively. The $i$-th element of $\textbf{x}$ and the $(i,j)$-th entry of $\textbf{X}$ are denoted by $[\textbf{x}]_i$ and $[\textbf{X}]_{i,j}$, respectively. The $\ell_2$-norm and $\ell_{\infty}$-norm of $\textbf{x}$ are denoted by $\lVert \textbf{x} \rVert$ and $\lVert \textbf{x} \rVert_{\infty}$, respectively, where $\lVert \cdot \rVert_{\infty}$ returns the maximum magnitude among the elements of its argument. The notation $\textbf{I}_D$ refers to the $D \times D$ identity matrix. The operator $\mathbb{E}\{\cdot\}$ denotes expectation, $\operatorname{Var}\{\cdot\}$ denotes variance, $\delta(\cdot)$ denotes the Dirac delta function, $\lfloor \cdot \rfloor$ denotes the floor function, $\Re\{\cdot\}$ returns the real part of its argument, and $\operatorname{diag}(\lambda_1,\dots,\lambda_N)$ denotes an $N \times N$ diagonal matrix with diagonal entries $\lambda_1,\dots,\lambda_N$ and zeros elsewhere.
%
%
%

\vspace{-2mm}
\section{SYSTEM MODEL} \label{sec_system_model}
This study focuses on the blind detection of a single target signal within the observed band. In this section, we first introduce the signal model in Section~\ref{sec_system_model}-\ref{sec_IQ_demod}. Subsequently, an asymptotic covariance analysis is conducted in Section~\ref{sec_system_model}-\ref{sec_asymptotical_PSD}, which enables the development of a discrete-time signal model. Finally, Section~\ref{sec_system_model}-\ref{sec_SNR_def} discusses the SNR definition from a physically meaningful perspective.

\vspace{-3mm}
\subsection{Received Signal Model} \label{sec_IQ_demod}
Suppose that the receiver is configured to monitor the frequency band from $f_{\text{min}}$ to $f_{\text{max}}$. Any signal activity outside this band is not of interest. Accordingly, we assume that the received signal lies entirely within this frequency range.
The received signal $y(t)$ at the receiver can be expressed in the time domain as
\begin{equation}
    y(t) = \Re \left\{\sqrt{P_{\textrm{s}}} x(t) e^{j2\pi f_c t} + \omega(t) \right\},
\end{equation}
for $0 \leq t \leq T$, where $T$ denotes the observation duration, $P_{\textrm{s}}$ denotes the power-scaling parameter, $x(t)$ is the normalized complex baseband representation of the transmitted OFDM signal with bandwidth $W$ such that the corresponding passband signal occupies the frequency range from $f_c - W/2$ to $f_c + W/2$, $f_c \in \mathcal{U}[f_{\text{min}} + W/2, f_{\text{max}} - W/2]$ denotes the carrier frequency, $\omega(t)$ represents circularly symmetric complex additive white Gaussian noise with autocorrelation $\mathbb{E}\{\omega(t_1)\omega^*(t_2)\} = N_0 \delta(t_1 - t_2)$, and $N_0$ is the noise power spectral density (W/Hz).
We assume that the signal power $P_{\textrm{s}}$, the bandwidth $W$, and the carrier frequency $f_c$ are unknown.

The analog continuous-time signal is first passed through a bandpass filter that allows the frequency range $[f_{\text{min}}, f_{\text{max}}]$. It is then downconverted to baseband via in-phase and quadrature (IQ) demodulation using a local oscillator with frequency $f_{\textrm{LO}} \triangleq (f_{\text{min}} + f_{\text{max}})/2$, followed by lowpass filtering to suppress the high-frequency components. The signal is then sampled at a rate $f_s \triangleq f_{\text{max}} - f_{\text{min}}$. Under ideal filtering assumptions, the resulting complex discrete-time signal can be expressed as
\begin{equation} \label{eq_discrete_model}
    y[n] = \sqrt{P_{\textrm{s}}} x[n] e^{j2\pi n \frac{\Delta f}{f_s}} + \omega[n],
\end{equation}
for $n = 1,2,\dots,N$, where $N \triangleq \lfloor T f_s \rfloor$ denotes the number of received samples, and $\Delta f \triangleq f_c - f_{\textrm{LO}}$ represents the residual frequency offset with $|\Delta f| \leq (f_s-W)/2$.
%
In the discrete-time model, the baseband signal samples are assumed to be normalized, i.e., $\mathbb{E}\{|x[n]|^2\} = 1$.
Under this normalization, $P_{\textrm{s}}$ represents the signal power. The complex noise samples $\omega[n]$ are independent and identically distributed (i.i.d.) according to $\mathcal{CN}(0,\sigma^2)$, where $\sigma^2 \triangleq  N_0 f_s$ denotes the noise sample power in watts.

Notably, the sampling frequency $f_s$ must be at least $f_{\text{max}} - f_{\text{min}}$ to avoid aliasing according to the Nyquist criterion. On the other hand, if $f_s$ is chosen larger than $f_{\text{max}} - f_{\text{min}}$, the resulting noise samples become correlated, since the noise is bandlimited by the initial bandpass filter. Therefore, we select $f_s = f_{\text{max}} - f_{\text{min}}$ to satisfy the Nyquist criterion while avoiding correlated noise samples.\footnote{In this study, increasing the sampling rate $f_s$ corresponds to enlarging the observed band, with $f_s \triangleq f_{\text{max}} - f_{\text{min}}$. By contrast, increasing $f_s$ while keeping the observed band fixed would produce correlated noise samples, which is outside the scope of this paper.}

\vspace{-3mm}
\subsection{Asymptotical Covariance Analysis} \label{sec_asymptotical_PSD}
\begin{figure}[ht]
    \vspace{3mm}
    \centering
    \includegraphics[width=1\columnwidth]{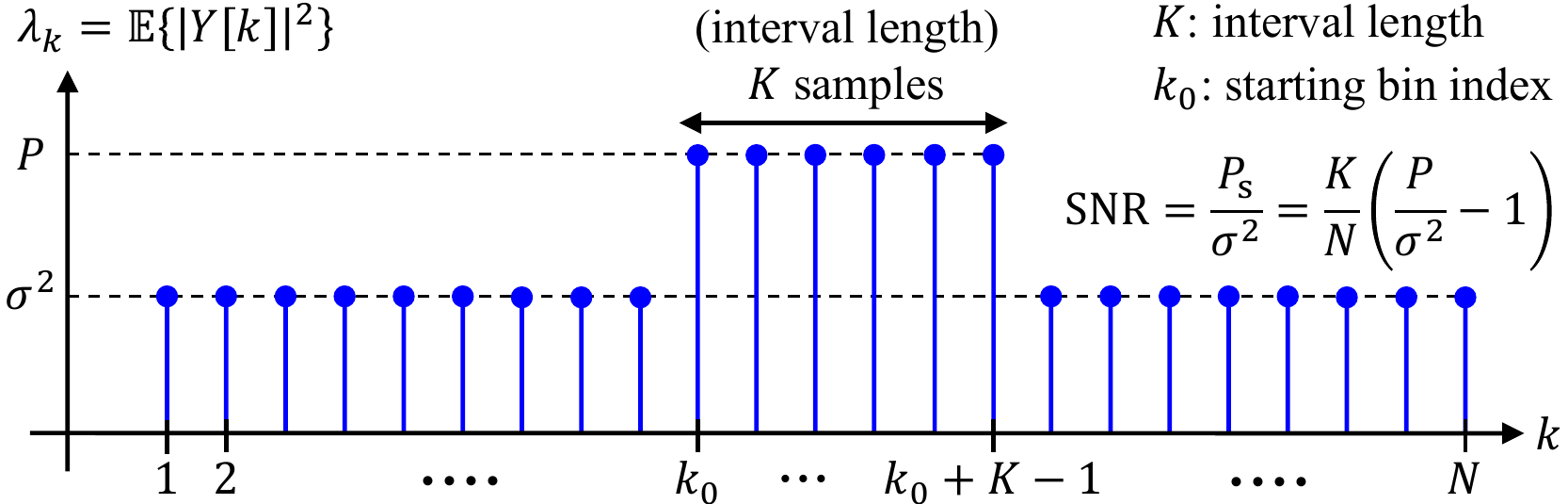}
    \caption{Illustration of DFT bin power levels}
    \label{fig_rect_PSD}
    \vspace{2mm}
\end{figure}

%
%
We define the received time-domain signal vector as
\begin{equation}
    \textbf{y} \triangleq \big[y[1] \; y[2] \; \dots \; y[N] \big]^T \in \mathbb{C}^{N \times 1}.
\end{equation}
Let $\textbf{R}_{\textbf{y}} \triangleq \mathbb{E}\{\textbf{y}\textbf{y}^H\} \in \mathbb{C}^{N \times N}$ denote the covariance matrix of the received signal $\textbf{y}$. Notably, $\textbf{R}_{\textbf{y}}$ is a Toeplitz matrix. As $N \to \infty$, Toeplitz covariance matrices asymptotically exhibit a circulant structure \cite{toeplitz,Szego}. Therefore, under the circulant approximation, the covariance matrix $\textbf{R}_{\textbf{y}}$ admits the following eigenvalue decomposition \cite{toeplitz,Szego}:
\begin{equation} \label{eq_circulant}
    \textbf{R}_{\textbf{y}} \approx \textbf{F}^H \boldsymbol{\Lambda} \textbf{F},
\end{equation}
where $\textbf{F} \in \mathbb{C}^{N \times N}$ denotes the unitary DFT matrix, and $\boldsymbol{\Lambda} = \operatorname{diag}\{\lambda_1,\lambda_2,\dots,\lambda_N\} \in \mathbb{C}^{N \times N}$ is a diagonal matrix whose diagonal entries correspond to the discrete PSD of the received signal. The DFT matrix is given by\footnote{Multiplication by $(-1)^{n-1}$ in \eqref{eq_DFT_matrix} corresponds to a frequency shift of $\pi$ in the discrete-time Fourier transform (DTFT), centering the spectrum around $[-f_s/2,f_s/2]$; the DFT samples then reflect this centered content.}
\begin{equation} \label{eq_DFT_matrix}
    [\textbf{F}]_{m,n} = \tfrac{1}{\sqrt{N}} (-1)^{(n-1)} e^{-j \frac{2\pi}{N}(m-1)(n-1)},
\end{equation}
for $m,n \in \{1,2,\dots,N\}$.
Importantly $\textbf{F} \textbf{F}^H = \textbf{I}_N$.
We define the DFT of the signal $\textbf{y}$ as
\begin{equation} \label{eq_DFT}
    \widetilde{\textbf{y}} = \big[Y[1] \; Y[2] \; \dots \; Y[N] \big]^T \triangleq \textbf{F} \textbf{y} \in \mathbb{C}^{N \times 1}.
\end{equation}
%
%
%
%
Using \eqref{eq_circulant} and \eqref{eq_DFT}, the frequency-domain covariance $\textbf{R}_{\widetilde{\textbf{y}}} \triangleq \mathbb{E}\{\widetilde{\textbf{y}} \widetilde{\textbf{y}}^H\}$ can be written as
\begin{equation} \label{eq_diagonal_CCM}
\begin{aligned}
    \textbf{R}_{\widetilde{\textbf{y}}} 
    = \mathbb{E}\{\textbf{F} \textbf{y} \textbf{y}^H \textbf{F} ^H\}
    = \textbf{F} \textbf{R}_{\textbf{y}} \textbf{F}^H
    \approx \textbf{F} (\textbf{F}^H \boldsymbol{\Lambda} \textbf{F}) \textbf{F}^H
    = \boldsymbol{\Lambda}.
\end{aligned}
\end{equation}
Equation~\eqref{eq_diagonal_CCM} shows that the frequency-domain covariance matrix $\textbf{R}_{\widetilde{\textbf{y}}}$ is asymptotically diagonal, implying that the DFT bins $Y[k]$'s are uncorrelated. Consequently, under the Gaussian assumption, the DFT bins are asymptotically independent.\footnote{The Gaussian assumption is well suited for OFDM signals, since the time-domain OFDM waveform is formed by the superposition of a large number of subcarriers, which, by the central limit theorem, approaches a Gaussian distribution.}
Based on this observation, as $N \to \infty$, the received time-domain signal in \eqref{eq_discrete_model} can be equivalently represented in the frequency domain as
\begin{equation} \label{eq_freq_domain_discrete_model}
    Y[k] \sim \mathcal{CN}(0,\lambda_k),
\end{equation}
for $k = 1,2,\dots,N$, where $Y[1], Y[2], \dots, Y[N]$ are mutually independent.
In the DFT domain, the frequency resolution is $1/T$. For a signal with bandwidth $W$, the number of occupied DFT bins, or equivalently, the interval length, is given by $K \triangleq \lfloor WT \rfloor$. We define the set of consecutive DFT bin indices containing the signal as
\begin{equation} \label{eq_signal_band}
    \mathcal{K} \triangleq \{k_0, k_0+1, \dots, k_0 + K-1 \},
\end{equation}
where $k_0$ denotes the starting index of the signal band.
Accordingly, the DFT bins with indices $k \notin \mathcal{K}$ contain only noise.
Under the rectangular PSD assumption\footnote{The rectangular PSD assumption is well justified for OFDM signals, as their spectra are approximately rectangular.}, the power levels in \eqref{eq_freq_domain_discrete_model} can be expressed as
\begin{equation} \label{eq_rect_PSD_approx}
    \lambda_k = P_{\textrm{s}}  \mathbb{E} \big\{ \big| X[k] \big|^2 \big\} + \sigma^2 = \begin{cases}
            \sigma^2 & k \notin \mathcal{K},\\[0.2em]
            P  & k \in \mathcal{K},
\end{cases}
\end{equation}
where $X[k]$ denotes the DFT of the time-domain signal $x[n] e^{j2\pi n \frac{\Delta f}{f_s}}$ in \eqref{eq_discrete_model}, and $P$ represents the power per DFT bin within the signal band.
For the remainder of this paper, we adopt the PSD model in \eqref{eq_rect_PSD_approx}, illustrated in Fig.~\ref{fig_rect_PSD}.

From Parseval's relation, we can write $\operatorname{Tr}\{\textbf{R}_{\textbf{y}}\} = \operatorname{Tr}\{\textbf{R}_{\widetilde{\textbf{y}}}\}$. From the time-domain signal model in \eqref{eq_discrete_model}, we have $\operatorname{Tr}\{\textbf{R}_{\textbf{y}}\} = N P_{\textrm{s}} \mathbb{E}\{|x[n]|^2\} + N\sigma^2 = N P_{\textrm{s}} + N\sigma^2$, since $\mathbb{E}\{|x[n]|^2\} = 1$. On the other hand, from the frequency-domain PSD model in \eqref{eq_rect_PSD_approx}, we have $\operatorname{Tr}\{\textbf{R}_{\widetilde{\textbf{y}}}\} = KP + (N-K)\sigma^2$. Equating these two expressions yields $N P_{\textrm{s}} + N\sigma^2 = KP + (N-K)\sigma^2$. Consequently, the in-band power level $P$ and the SNR can be expressed as
\begin{equation}
    P = \frac{N}{K} P_{\textrm{s}} + \sigma^2 \quad , \quad \textrm{SNR} \triangleq \frac{P_{\textrm{s}}}{\sigma^2} = \frac{K}{N} \Big( \frac{P}{\sigma^2}-1 \Big).
\end{equation}

Table~\ref{table_params} summarizes the relationships between the discrete-time parameters and physical parameters.

\vspace{2mm}
\begin{table}[ht]
\caption{Discrete-time signal parameters}
\label{table_params}
\centering
\setlength\tabcolsep{4pt}
\renewcommand{\arraystretch}{1.8}
\resizebox{0.95\columnwidth}{!}{%
\begin{tabular}{c|c}
\hline
\multicolumn{1}{c|}{Discrete-time parameters} & \multicolumn{1}{c}{Relation to physical parameters} \\ \hline
No. of received samples $(N)$                                                            & $N = \lfloor T f_s \rfloor$                                                                                        \\ \hline
No. of in-band DFT bins $(K)$                                                           & $K = \lfloor T W \rfloor$\\ \hline
 Starting DFT bin index $(k_0)$&$k_0 = \lfloor N(\frac{\Delta f}{f_s} - \frac{W}{2f_s} + \frac{1}{2}) \rfloor + 1$\\ \hline
Noise bin power $(\sigma^2)$                                                                & $\sigma^2 = N_0 f_s$                                                                                        \\ \hline
In-band bin power $(P)$& $P = \frac{N}{K} P_{\textrm{s}} + \sigma^2$\\ \hline
Signal-to-noise ratio (SNR)                                                             & $\textrm{SNR} = {P_{\textrm{s}}}/{\sigma^2} $ \\ \hline
\end{tabular}
}
\end{table}

\vspace{-3mm}
\subsection{Discussion on the SNR Definition} \label{sec_SNR_def}
The definition of SNR in blind spectrum sensing is subtle and requires careful consideration. In \cite{GLRT_paper}, SNR is defined as $(P/\sigma^2 - 1)$, with $P$ and $\sigma^2$ illustrated in Fig.~\ref{fig_rect_PSD}. A similar approach is used in \cite{freq_hop}, where SNR is defined over the signal band rather than the entire observed band.
Under these definitions, keeping SNR fixed while increasing the signal bandwidth $W$ implicitly increases the total received signal power. For fair comparisons, the total signal power $P_{\textrm{s}}$ should instead remain constant. Accordingly, SNR can be defined as $\textrm{SNR} \triangleq P_{\textrm{s}}/\sigma^2 = P_{\textrm{s}}/(N_0 f_s)$.
However, when comparing detectors that operate at different sampling rates, a fair comparison should be based on keeping the ratio $P_{\textrm{s}}/N_0$ fixed rather than keeping the SNR fixed.


\vspace{-2mm}
\section{PROPOSED LOW-COMPLEXITY DETECTOR DESIGNS} \label{sec_detector_designs}
The received signal in the frequency domain can be modeled under the no-signal hypothesis $\mathcal{H}_0$ and the signal-present hypothesis $\mathcal{H}_1$ using \eqref{eq_freq_domain_discrete_model} as
\begin{equation} \label{eq_hypothesis_model}
    Y[k] \sim \begin{cases}
            \mathcal{CN}(0,\sigma^2) & \mathcal{H}_0,\\[0.2em]
            \mathcal{CN}(0,\lambda_k)  & \mathcal{H}_1,
\end{cases}
\end{equation}
for $k = 1,2,\dots,N$, where $\lambda_k$ is given in \eqref{eq_rect_PSD_approx}.
Notably, the set $\mathcal{K}$ in \eqref{eq_signal_band}, parameterized by $k_0$ and $K$, as well as the per-bin power within the signal band, are unknown; that is, $k_0$, $K$, and $P$ are unknown.

\vspace{-4mm}
\subsection{GLRT Detector with Known Noise Power} \label{sec_GLRT_known_noise}
The GLRT metric for known noise power $\sigma^2$ can be expressed as
\begin{equation} \label{eq_GLRT_metric}
    \widetilde{\mathcal{L}}_1 = \frac{ \underset{\mathcal{K},P}{\text{max} } \hspace{1mm} p(\widetilde{\textbf{y}} \hspace{0.5mm} | \hspace{0.5mm} \mathcal{H}_1)}{p(\widetilde{\textbf{y}} \hspace{0.5mm} | \hspace{0.5mm} \mathcal{H}_0)}.
\end{equation}
The likelihoods in \eqref{eq_GLRT_metric} can be calculated using \eqref{eq_hypothesis_model} as
\begin{equation} \label{eq_likelihood_H0}
    p(\widetilde{\textbf{y}} \hspace{0.5mm} | \hspace{0.5mm} \mathcal{H}_0)
    =
    \prod_{k=1}^{N} \frac{1}{\pi \sigma^2} \operatorname{exp} \left\{-{|Y[k]|^2}/{\sigma^2} \right\},
\end{equation}
\begin{equation} \label{eq_likelihood_H1}
    p(\widetilde{\textbf{y}} \hspace{0.5mm} | \hspace{0.5mm} \mathcal{H}_1)
    =
    \prod_{k=1}^{N} \frac{1}{\pi \lambda_k} \operatorname{exp} \left\{-{|Y[k]|^2}/{\lambda_k} \right\}.
\end{equation}
Substituting \eqref{eq_likelihood_H0} and \eqref{eq_likelihood_H1} into \eqref{eq_GLRT_metric}, and using the log-likelihood function, the GLRT can be expressed as
\begin{equation} \label{eq_GLRT_metric_log}
    \mathcal{L}_1 = N \ln{\sigma^2} + \underset{\mathcal{K},P}{\text{max}} \hspace{1mm} \sum_{k=1}^N \left( \frac{|Y[k]|^2}{\sigma^2} -\frac{|Y[k]|^2}{\lambda_k} - \ln{\lambda_k} \right).
\end{equation}
Using \eqref{eq_rect_PSD_approx}, the GLRT in \eqref{eq_GLRT_metric_log} can be written as
\begin{equation} \label{eq_GLRT_metric_log_simplified}
    \mathcal{L}_1 = \underset{\mathcal{K},P}{\text{max}} \hspace{1mm}  \Big(\frac{1}{\sigma^2} - \frac{1}{P} \Big) \Big(\sum_{k \in \mathcal{K}}  |Y[k]|^2 \Big) - K \ln{\frac{P}{\sigma^2}} .
\end{equation}
From the derivative test, the optimal value of $P$, for the feasible region\footnote{The feasible region consists of values for which $\widehat{P} \geq \sigma^2$ in \eqref{eq_P_hat}. If this condition is not satisfied for any $\mathcal{K}$, the detector decides $\mathcal{H}_0$.}, is obtained as
\begin{equation} \label{eq_P_hat}
    \widehat{P} = \frac{1}{K}  \sum_{k \in \mathcal{K}}  |Y[k]|^2.
\end{equation}
Substituting \eqref{eq_P_hat} into \eqref{eq_GLRT_metric_log_simplified}, we obtain
%
%
\begin{equation} \label{eq_GLRT_metric_final}
    \mathcal{L}_1 = \underset{k_0, K}{\text{max} } \hspace{1mm} K \Big(\frac{\widehat{P}}{\sigma^2} - \ln{\frac{\widehat{P}}{\sigma^2}} - 1 \Big).
\end{equation}
Notably, for a given $K$, the GLRT metric in \eqref{eq_GLRT_metric_final} is a monotonically increasing function of $\widehat{P}$ for $\widehat{P} \geq \sigma^2$, which is the regime of interest. Thus, for a given $K$, the problem reduces to finding the maximum value of $\widehat{P}$, i.e., the $K$-sample subband with the highest power.
To this end, we define the CUSUM vector $\textbf{c} \in \mathbb{R}^{(N+1) \times 1}$ as
\begin{equation} \label{eq_cusum}
    [\textbf{c}]_m = \begin{cases}
            0 & m=1,\\[0.2em]
            \sum_{k=1}^{m-1} |Y[k]|^2  & m \in \{2,3,\dots,N+1 \}.
\end{cases}
\end{equation}
Notably, the difference vector $[\textbf{c}]_{(K+1:N+1)} - [\textbf{c}]_{(1:N-K+1)}$ yields the powers of all $K$-sample subbands. 
Accordingly, for a given $K$, we define
\begin{equation} \label{eq_opt_P_K}
    P_K \triangleq \frac{1}{K} \big \lVert [\textbf{c}]_{(K+1:N+1)} - [\textbf{c}]_{(1:N-K+1)} \big \rVert_{\infty},
\end{equation}
which corresponds to the average power of the $K$-sample subband with the highest average power. Then, the GLRT in \eqref{eq_GLRT_metric_final} can be expressed in its final form as
\begin{equation} \label{eq_GLRT_metric_final_compact}
    \mathcal{L}_1 = \underset{K}{\text{max} } \hspace{1mm} K \Big(\frac{P_K}{\sigma^2} - \ln{ \frac{P_K}{\sigma^2}} - 1 \Big) .
\end{equation}
%
%
The implementation of the GLRT with known noise power is summarized in Algorithm~\ref{Algo_1}.
Importantly, to reduce computational complexity, we perform a dyadic search over the interval length $K$, considering candidate values from the set $\mathcal{S}_K = \{1, 2, 4, \dots, 2^{\lfloor \log_2 N \rfloor}\}$. This approach is shown to be near-optimal in \cite{GLRT_paper} and \cite{dyadic_1}.

\setlength{\textfloatsep}{2pt}
\begin{algorithm}
{\fontsize{10pt}{10pt}\selectfont
\footnotesize
\caption{GLRT with known noise power}\label{Algo_1}
\begin{algorithmic}[1]
\Require $Y[k]$ for $k=1,\dots,N$ , Noise power: $\sigma^2$, Threshold: $\gamma_1$


    \vspace{1mm}
    \State Calculate the CUSUM in \eqref{eq_cusum}: $\textbf{c} \in \mathbb{R}^{(N+1) \times 1}$
    
    %

    \vspace{1mm}
    \For{$K \in \mathcal{S}_K = \{1,2,4,8,\dots,2^{\lfloor \log_2{N} \rfloor} \}$}
        \vspace{1mm}
        \State $P_K = \operatorname{max} \left\{ \frac{1}{K} \big \lVert [\textbf{c}]_{(K+1:N+1)} - [\textbf{c}]_{(1:N-K+1)} \big \rVert_{\infty}, \sigma^2 \right\}$

        \vspace{1mm}
        \State $\mathcal{L}_1(K) = K \Big(\frac{P_K}{\sigma^2} - \ln{\frac{P_K}{\sigma^2}} - 1 \Big)$
        %
    
    \vspace{1mm}
    \EndFor 

    \vspace{1mm}
    \State Decision: $\mathcal{H}_1$ if $\max_{K \in \mathcal{S}_K} \{ \mathcal{L}_1(K) \} \geq \gamma_1$, and $\mathcal{H}_0$ otherwise

\vspace{1mm}    
\Output Decision: $\mathcal{H}_0 \textrm{ or } \mathcal{H}_1$
\end{algorithmic}}
\end{algorithm}

\subsection{GLRT Detector with Unknown Noise Power} \label{sec_GLRT_unknown_noise}
The GLRT metric for unknown noise power $\sigma^2$ can be expressed as
\begin{equation} \label{eq_GLRT2_metric}
    \widetilde{\mathcal{L}}_2 = \frac{ \underset{\mathcal{K},P,\sigma_1^2}{\text{max} } \hspace{1mm} p(\widetilde{\textbf{y}} \hspace{0.5mm} | \hspace{0.5mm} \mathcal{H}_1)}{ \underset{\sigma_0^2}{\text{max} } \hspace{1mm} p(\widetilde{\textbf{y}} \hspace{0.5mm} | \hspace{0.5mm} \mathcal{H}_0)}.
\end{equation}
In \eqref{eq_GLRT2_metric}, the noise powers are denoted by $\sigma_1^2$ in the numerator and $\sigma_0^2$ in the denominator to avoid notational confusion.
%
%
With the log-likelihood function, the numerator can be calculated using \eqref{eq_rect_PSD_approx} and \eqref{eq_likelihood_H1} as
\begin{equation} \label{eq_numerator}
    \mathcal{L}_{\textrm{n}} = - (N-K) \ln{\sigma_1^2} - K \ln{P} - \sum_{k \notin \mathcal{K}} \frac{|Y[k]|^2}{\sigma_1^2} - \sum_{k \in \mathcal{K}} \frac{|Y[k]|^2}{P}.
\end{equation}
From the derivative test, the optimal values of $P$ and $\sigma_1^2$ are obtained, for the feasible region, as
\begin{equation} \label{eq_P2_hat}
    \widehat{P} = \frac{1}{K} \sum_{k \in \mathcal{K}} |Y[k]|^2,
\end{equation}
\begin{equation} \label{eq_sigma_1_hat}
    \widehat{\sigma}_1^2 = \frac{1}{N-K} \sum_{k \notin \mathcal{K}} |Y[k]|^2.
\end{equation}
Substituting \eqref{eq_P2_hat} and \eqref{eq_sigma_1_hat} into \eqref{eq_numerator}, we obtain
\begin{equation} \label{eq_numerator_2}
    \mathcal{L}_{\textrm{n}} = \underset{\mathcal{K}}{\text{max} } \hspace{1mm} \Big(-(N-K) \ln{\widehat{\sigma}_1^2} - K \ln{\widehat{P}} -N \Big).
\end{equation}
Similarly, the denominator can be calculated by \eqref{eq_likelihood_H0} as
\begin{equation} \label{eq_denominator}
    \mathcal{L}_{\textrm{d}} = {\underset{\sigma_0^2}{\text{max} }} \hspace{1mm} \Big( -N \ln{\sigma_0^2} - \frac{1}{{\sigma_0^2}}\sum_{k=1}^N |Y[k]|^2 \Big ).
\end{equation}
From the derivative test, the optimal value of $\sigma_0^2$ is obtained as
\begin{equation} \label{eq_sigma_0_hat}
    \widehat{\sigma}_0^2 = \frac{1}{N} \sum_{k=1}^N |Y[k]|^2.
\end{equation}
Note that \eqref{eq_P2_hat}, \eqref{eq_sigma_1_hat}, and \eqref{eq_sigma_0_hat} together imply that $\widehat{\sigma}_1^2 = \frac{N \widehat{\sigma}_0^2 - K \widehat{P}}{N-K}$.
Plugging \eqref{eq_sigma_0_hat} into \eqref{eq_denominator}, we obtain
\begin{equation} \label{eq_denominator_2}
    \mathcal{L}_{\textrm{d}} = -N \ln{\widehat{\sigma}_0^2} - N.
\end{equation}
Combining the calculations of the numerator and denominator in \eqref{eq_numerator_2} and \eqref{eq_denominator_2}, respectively, the GLRT $\mathcal{L}_2 = \mathcal{L}_{\textrm{n}} - \mathcal{L}_{\textrm{d}}$ can be written as
%
%
%
\begin{equation} \label{eq_GLRT2_metric_final}
    \mathcal{L}_2 = \underset{k_0,K}{\text{max} } \hspace{1mm} \Big( N \ln{\widehat{\sigma}_0^2} -(N-K) \ln{\widehat{\sigma}_1^2} - K \ln{\widehat{P}} \Big).
\end{equation}
Similar to \eqref{eq_GLRT_metric_final}, the GLRT metric increases monotonically with $\widehat{P}$ for a given $K$. Consequently, the GLRT can be expressed in its final form as
%
%
%
\begin{equation} \label{eq_GLRT2_metric_final_compact}
    \mathcal{L}_2 =  \underset{K}{\text{max} } \Big(N \ln{\widehat{\sigma}_0^2}-(N-K) \ln{\widehat{\sigma}_1^2} - K \ln{P_K} \Big),
\end{equation}
where $P_K$ is given in \eqref{eq_opt_P_K} and $\widehat{\sigma}_1^2 = \frac{N \widehat{\sigma}_0^2 - K P_K}{N-K}$.
%
%
The corresponding implementation is given in Algorithm~\ref{Algo_2}.

\setlength{\textfloatsep}{20pt}
\begin{algorithm}
{\fontsize{10pt}{10pt}\selectfont
\footnotesize
\caption{GLRT with unknown noise power}\label{Algo_2}
\begin{algorithmic}[H]
\Require $Y[k]$ for $k=1,2,\dots,N$, Threshold: $\gamma_2$


    \vspace{1mm}
    \State Calculate the CUSUM in \eqref{eq_cusum}: $\textbf{c} \in \mathbb{R}^{(N+1) \times 1}$
    
    %

    \vspace{1mm}
    \State Calculate the average received power: $\widehat{\sigma}_0^2 = \frac{1}{N} \sum_{k=1}^N |Y[k]|^2$
        
    \vspace{1mm}
    \For{$K \in \mathcal{S}_K = \{1,2,4,8,\dots,2^{\lfloor \log_2{N} \rfloor} \}$}
        \vspace{1mm}
        \State $P_K = \frac{1}{K} \big \lVert [\textbf{c}]_{(K+1:N+1)} - [\textbf{c}]_{(1:N-K+1)} \big \rVert_{\infty}$

        \vspace{1mm}
        \State $\widehat{\sigma}_1^2 = \frac{N \widehat{\sigma}_0^2 - K P_K}{N-K}$
        
        \vspace{1mm}
        \State $\mathcal{L}_2(K) = N \ln{\widehat{\sigma}_0^2} -(N-K) \ln{\widehat{\sigma}_1^2} - K \ln{P_K}$
    
    \vspace{1mm}
    \EndFor 

    \vspace{1mm}
    \State Decision: $\mathcal{H}_1$ if $\max_{K \in \mathcal{S}_K} \{ \mathcal{L}_2(K) \} \geq \gamma_2$, and $\mathcal{H}_0$ otherwise

\vspace{1mm}    
\Output Decision: $\mathcal{H}_0 \textrm{ or } \mathcal{H}_1$
\end{algorithmic}}
\end{algorithm}

\subsection{Discussion on the CUSUM} \label{sec_cusum}
\begin{figure}[ht]
    \vspace{3mm}
    \centering
    \includegraphics[width=0.99\columnwidth]{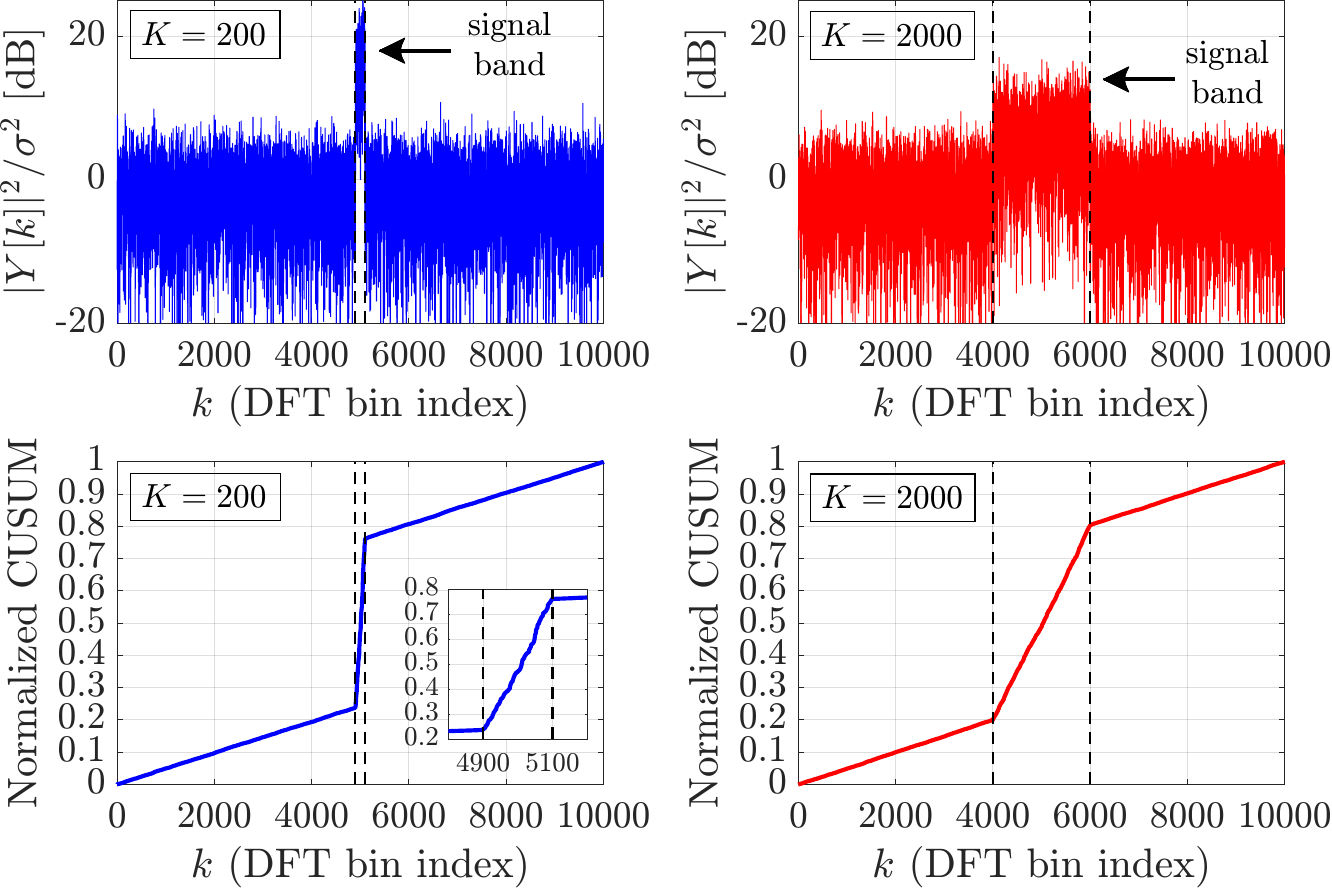}
    \vspace{0.1mm}
    \caption{DFT-bin power levels and the corresponding normalized CUSUM}
    \label{fig_cusum}
    \vspace{2mm}
\end{figure}

%
%
This subsection provides intuition for the role of the CUSUM in the proposed GLRT detectors. As shown in Algorithms \ref{Algo_1} and \ref{Algo_2}, the CUSUM constitutes a sufficient statistic for the proposed GLRT detectors.
Fig.~\ref{fig_cusum} illustrates the squared magnitudes of the DFT bins alongside the corresponding normalized CUSUM, with $N=10^4$ and $\textrm{SNR}=0$ dB.
It can be observed that a smaller interval length $K$, corresponding to a narrower signal bandwidth, results in a higher spectral density level. This behavior is also reflected in the CUSUM, where the slope change is more pronounced for smaller values of $K$. Intuitively, this explains why, for a fixed received signal power, spread signals are generally more difficult to detect. This relationship between detectability and signal bandwidth is further quantified in Table \ref{table_impact_of_params} in Section \ref{sec_theoretical_analysis} and supported by the simulation results.

Owing to its normalized structure bounded between $0$ and $1$, and its smooth behavior compared to the highly fluctuating DFT-bin samples, the CUSUM is well suited for potential ML/DL-based detection frameworks. While this study focuses on a traditional detection framework, using CUSUM in ML/DL framework for multi-target detection represents a promising direction for future work.

\section{EXTENSION TO MULTI-ANTENNA CASE} \label{sec_multiantenna_extension}
%
%
Suppose the receiver is equipped with a uniform linear array (ULA) consisting of $M$ antennas. The received signal can then be expressed in the frequency domain as
\begin{equation} \label{eq_multi_antenna_signal_model}
    \textbf{y}_k = \begin{cases}
            \boldsymbol{w}_k \in \mathbb{C}^{M \times 1} & \mathcal{H}_0,\\[0.2em]
            \sqrt{P_{\textrm{s}}} \textbf{u}(\theta) X[k]  + \boldsymbol{w}_k \in \mathbb{C}^{M \times 1}  & \mathcal{H}_1,
\end{cases}
\end{equation}
for $k = 1,2,\dots,N$, where $\textbf{u}(\cdot) \in \mathbb{C}^{M \times 1}$ denotes the array response vector, $\theta \triangleq \sin(\widetilde{\theta})$ with $\widetilde{\theta}$ being the physical angle of arrival (AoA), and $\boldsymbol{w}_k \sim \mathcal{CN}(0,\sigma^2\textbf{I}_M)$ denotes the noise vector. The array response vector is given by
\begin{equation}
    \textbf{u}(\theta) = [1 \; e^{-j2\pi \frac{d}{\lambda} \theta} \; \dots \; e^{-j2\pi (M-1) \frac{d}{\lambda} \theta}]^T,
\end{equation}
where $d$ is the antenna spacing and $\lambda$ is the wavelength of the signal. 
%
%
We define power levels, similar to \eqref{eq_rect_PSD_approx}, as
\begin{equation} \label{eq_multi_antenna_power_levels}
    \lambda_{x_k} \triangleq P_{\textrm{s}} M\mathbb{E}\{ |X[k] |^2 \} + \sigma^2  = \begin{cases}
            \sigma^2 & k \notin \mathcal{K},\\[0.2em]
            P_y  & k \in \mathcal{K}.
\end{cases}
\end{equation}
%
%
The received signal can then be modeled as
\begin{equation}
    \textbf{y}_k \sim \begin{cases}
            \mathcal{CN}(0,\sigma^2\textbf{I}_M) & \mathcal{H}_0,\\[0.2em]
            \mathcal{CN}(0,\textbf{R}_{\textbf{y}_k})  &  \mathcal{H}_1,
\end{cases}
\end{equation}
where $\textbf{R}_{\textbf{y}_k} \triangleq \mathbb{E}\{\textbf{y}_k \textbf{y}^H_k \mid \mathcal{H}_1\} \in \mathbb{C}^{M \times M}$. This covariance 

\noindent can be calculated using \eqref{eq_multi_antenna_signal_model} and \eqref{eq_multi_antenna_power_levels} as
\begin{equation}
    \textbf{R}_{\textbf{y}_k} =  \frac{1}{M} (\lambda_{x_k} - \sigma^2) \textbf{u}(\theta)\textbf{u}^H(\theta) + \sigma^2\textbf{I}_M.
\end{equation}
The log-likelihoods can then be expressed as
\begin{equation} \label{eq_multi_antenna_likelihood_H0}
    \ln{p(\textbf{y}_k | \mathcal{H}_0)} = -\ln{\big(\pi^M (\sigma^2)^{M}\big)} - \frac{1}{\sigma^2} \textbf{y}_k^H \textbf{y}_k,
\end{equation}
\begin{equation} \label{eq_multi_antenna_likelihood_H1}
    \ln{p(\textbf{y}_k | \mathcal{H}_1)} = -\ln{\big(\pi^M \operatorname{det}|\textbf{R}_{\textbf{y}_k}|\big)} - \textbf{y}_k^H \textbf{R}_{\textbf{y}_k}^{-1} \textbf{y}_k.
\end{equation}
Notably, the determinant and the inverse of $\textbf{R}_{\textbf{y}_k}$ can be calculated using the Sherman–Morrison formula as
\begin{equation} \label{eq_det_Q}
    \operatorname{det}|\textbf{R}_{\textbf{y}_k}| = \lambda_{x_k} (\sigma^2)^{M-1},
\end{equation}
\begin{equation} \label{eq_O_inv}
    \textbf{R}_{\textbf{y}_k}^{-1} = \frac{1}{\sigma^2} \textbf{I}_M -  \Big(\frac{1}{\sigma^2} - \frac{1}{\lambda_{x_k}} \Big) \frac{1}{M} \textbf{u}(\theta)\textbf{u}^H(\theta).
\end{equation}
Substituting \eqref{eq_det_Q} and \eqref{eq_O_inv} into \eqref{eq_multi_antenna_likelihood_H1}, we obtain
\begin{equation}\label{eq_multi_antenna_likelihood_H1_v2}
\begin{aligned}
    \ln{p(\textbf{y}_k \hspace{0.5mm} | \hspace{0.5mm} \mathcal{H}_1)} = &-\ln{\big(\pi^M \lambda_{x_k} (\sigma^2)^{M-1} \big)} - \frac{1}{\sigma^2} \textbf{y}_k^H \textbf{y}_k
    \\&
    + \Big(\frac{1}{\sigma^2} - \frac{1}{\lambda_{x_k}} \Big) \frac{1}{M} |\textbf{u}^H(\theta) \textbf{y}_k|^2.
\end{aligned}
\end{equation}
Due to the independence of the $X[k]$’s, the joint likelihood is the product of the individual likelihoods over the DFT bins. Consequently, using \eqref{eq_multi_antenna_likelihood_H0} and \eqref{eq_multi_antenna_likelihood_H1_v2}, the joint likelihood under each hypothesis is given by
\begin{equation} \label{eq_multi_antenna_likelihood_H0_final}
    \ln{p\left(\{\textbf{y}_k\}_{k=1}^N  |  \mathcal{H}_0 \right)} = -NM \ln{\pi} - NM\ln{\sigma^2} - \sum_{k=1}^N \frac{\lVert \textbf{y}_k \rVert^2}{\sigma^2},
\end{equation}
\begin{equation} \label{eq_multi_antenna_likelihood_H1_final}
\begin{aligned}
    &\ln{p\left(\{\textbf{y}_k\}_{k=1}^N \hspace{0.5mm} | \hspace{0.5mm} \mathcal{H}_1 \right)} = -NM \ln{\pi} -(NM-K)\ln{\sigma^2} 
    \\&
    -K\ln{P_y}
    - \sum_{k=1}^N \frac{\lVert \textbf{y}_k \rVert^2}{\sigma^2}
    + \sum_{k \in \mathcal{K}} \Big(\frac{1}{\sigma^2} - \frac{1}{P_y} \Big) \frac{|\textbf{u}^H(\theta) \textbf{y}_k|^2}{M} .
\end{aligned}
\end{equation}

\vspace{-7mm}
\subsection{Multi-antenna GLRT with Known Noise Power} \label{sec_multi_antenna_GLRT_known_noise}
The GLRT with known noise power $\sigma^2$ can be expressed as
\begin{equation} \label{eq_multi_antenna_GLRT_metric}
    \widetilde{\mathcal{L}}_3 = \frac{ \underset{\mathcal{K},P_y,\theta}{\text{max} } \hspace{1mm} p \left(\{\textbf{y}_k\}_{k=1}^N \hspace{0.5mm} | \hspace{0.5mm} \mathcal{H}_1 \right)}{p \left(\{\textbf{y}_k\}_{k=1}^N \hspace{0.5mm} | \hspace{0.5mm} \mathcal{H}_0 \right)}.
\end{equation}
Substituting \eqref{eq_multi_antenna_likelihood_H0_final} and \eqref{eq_multi_antenna_likelihood_H1_final} into \eqref{eq_multi_antenna_GLRT_metric}, the GLRT can be expressed using the log-likelihood function as
\begin{equation} \label{eq_multi_antenna_GLRT_known_noise}
    \mathcal{L}_3 = \underset{\mathcal{K},P_y,\theta}{\text{max} } \bigg( \Big(\frac{1}{\sigma^2} - \frac{1}{P_y} \Big) \sum_{k \in \mathcal{K}} \frac{1}{M} |\textbf{u}^H(\theta) \textbf{y}_k|^2 -K\ln{\frac{P_y}{\sigma^2}} \bigg ).
\end{equation}
From the derivative test, the optimal value of $P_y$ can be obtained, for the feasible region, as
\begin{equation} \label{eq_Py_hat}
    \widehat{P}_y = \frac{1}{K} \sum_{k \in \mathcal{K}} \frac{1}{M} |\textbf{u}^H(\theta) \textbf{y}_k|^2.
\end{equation}
Substituting this into \eqref{eq_multi_antenna_GLRT_known_noise}, the GLRT simplifies to
\begin{equation}
    \mathcal{L}_3 = \underset{\mathcal{K},\theta}{\text{max} } \hspace{1mm} K \Big(\frac{\widehat{P}_y}{\sigma^2} - \ln{\frac{\widehat{P}_y}{\sigma^2}} - 1 \Big).
\end{equation}
Notably, for a given $K$, the GLRT metric is a monotonically increasing function of $\widehat{P}_y$ for $\widehat{P}_y \geq \sigma^2$, which is the regime of interest. Thus, for a fixed $K$, the problem reduces to finding the maximum value of $\widehat{P}_y$. However, the evaluation of $\widehat{P}_y$ in \eqref{eq_Py_hat} depends on the steering vector $\textbf{u}(\theta)$, which in turn depends on both the AoA $\theta$ and the carrier frequency $f_c$, leading to a 2D search.
This dependency can be eliminated by observing that the term $\frac{1}{M} | \textbf{u}^H(\theta)\textbf{y}_k |^2$ in \eqref{eq_Py_hat} corresponds to the squared magnitude of the spatial DFT operation applied to $\textbf{y}_k$. Accordingly, we define the matrix $\textbf{P} \in \mathbb{R}^{M \times N}$ as
\begin{equation} \label{eq_spatial_DFT}
    [\textbf{P}]_{m,k} \triangleq \big|[\textbf{F}_M \textbf{Y}]_{m,k} \big|^2 \in \mathbb{R},
\end{equation}
for $m=1,2,\dots,M$, $k=1,2,\dots,N$, where $\textbf{F}_M \in \mathbb{C}^{M \times M}$ denotes the $M \times M$ DFT matrix, which performs the spatial DFT across the antenna dimension, and $\textbf{Y}$ is the concatenated received signal matrix defined as
\begin{equation} \label{eq_concat_rx}
    \textbf{Y} \triangleq [\textbf{y}_1 \; \textbf{y}_2 \; \dots \; \textbf{y}_N] \in \mathbb{C}^{M \times N}.
\end{equation}
We then define the CUSUM matrix $\textbf{C} \in \mathbb{R}^{M \times (N+1)}$ as
\begin{equation} \label{eq_multi_antenna_cusum}
    [\textbf{C}]_{m,n} \triangleq \begin{cases}
    0 & n=1,
    \\[0.2em]
    \sum_{k=1}^{n-1} [\textbf{P}]_{m,k} & n \in \{2,3,\dots,N+1 \},\end{cases}
\end{equation}
for $m=1,2,\dots,M$.
For a given $K$, we then define 
\begin{equation} \label{eq_multi_antenna_optimal_P}
    P_K \triangleq \tfrac{1}{K} \lVert \textbf{C}_{(:,K+1:N+1)} - \textbf{C}_{(:,1:N-K+1)} \rVert_{\infty}.
\end{equation}
Importantly, $P_K$ represents the maximum accumulated power over all length-$K$ contiguous frequency segments and across all spatial DFT bins.
As a result, the GLRT can be expressed in its final form as
\begin{equation}
    \mathcal{L}_3 = \underset{K}{\text{max} } \hspace{1mm} K \Big(\frac{P_K}{\sigma^2} - \ln{\frac{P_K}{\sigma^2}} - 1 \Big).
\end{equation}
The proposed multi-antenna GLRT with known noise power is summarized in Algorithm~\ref{Algo_3}.
Importantly, despite the unknown parameters $k_0$, $K$, $P$, and $\theta$, Algorithm~\ref{Algo_3} requires only an effective 1D search.
%

\setlength{\textfloatsep}{2pt}
\begin{algorithm}[H]
{\fontsize{10pt}{10pt}\selectfont
\footnotesize
\caption{Multi-antenna GLRT with known noise power}\label{Algo_3}
\begin{algorithmic}[1]
\Require $\textbf{y}_k$ for $k=1,\dots,N$ in \eqref{eq_multi_antenna_signal_model} , Noise power: $\sigma^2$, Threshold: $\gamma_3$


    
    %
    
    \vspace{1mm}
    \State Calculate the CUSUM in \eqref{eq_multi_antenna_cusum}: $\textbf{C} \in \mathbb{R}^{M \times (N+1)}$
    
    %
    
    \vspace{1mm}
    \For{$K \in \mathcal{S}_K = \{1,2,4,8,16,\dots,2^{\lfloor \log_2{N} \rfloor} \}$}
        \vspace{1mm}
        \State $P_K = \operatorname{max} \left \{\frac{1}{K} \lVert \textbf{C}_{(:,K+1:N+1)} - \textbf{C}_{(:,1:N-K+1)} \rVert_{\infty}, \sigma^2 \right \}$

        \vspace{1mm}
        \State $\mathcal{L}_3(K) = K \Big(\frac{P_K}{\sigma^2} - \ln{\frac{P_K}{\sigma^2}} - 1 \Big)$
        %
    
    \vspace{1mm}
    \EndFor 

    \vspace{1mm}
    \State Decision: $\mathcal{H}_1$ if $\max_{K \in \mathcal{S}_K} \{ \mathcal{L}_3(K) \} \geq \gamma_3$, and $\mathcal{H}_0$ otherwise
    
\vspace{1mm}    
\Output Decision: $\mathcal{H}_0 \textrm{ or } \mathcal{H}_1$
\end{algorithmic}}
\end{algorithm}

\vspace{-7mm}
\subsection{Multi-antenna GLRT with Unknown Noise Power} \label{sec_multi_antenna_GLRT_unknown_noise}
The GLRT with unknown $\sigma^2$ can be expressed as
\begin{equation} \label{eq_multi_antenna_GLRT2_metric}
    \widetilde{\mathcal{L}}_4 = \frac{ \underset{\mathcal{K},\theta,P_y,\sigma_1^2}{\text{max} } \hspace{1mm} p\left(\{\textbf{y}_k\}_{k=1}^N \hspace{0.5mm} | \hspace{0.5mm} \mathcal{H}_1 \right)}{\underset{\sigma_0^2}{\text{max} } \hspace{1mm} p\left(\{\textbf{y}_k\}_{k=1}^N \hspace{0.5mm} | \hspace{0.5mm} \mathcal{H}_0 \right)}.
\end{equation}
With the log-likelihood function, the numerator can be calculated using \eqref{eq_multi_antenna_likelihood_H1_final} as
\begin{equation} \label{eq_multi_antenna_numerator}
\begin{aligned}
    &\mathcal{L}_{\textrm{n}} = -NM \ln{\pi} -(NM-K)\ln{\sigma^2} 
    \\&
    -K\ln{P_y}
    - \sum_{k=1}^N \frac{\lVert \textbf{y}_k \rVert^2}{\sigma^2}
    + \sum_{k \in \mathcal{K}} \Big(\frac{1}{\sigma^2} - \frac{1}{P_y} \Big) \frac{1}{M} |\textbf{u}^H(\theta) \textbf{y}_k|^2.
\end{aligned}
\end{equation}
From the derivative test, the optimal values of $P_y$ and $\sigma_1^2$ can be obtained, for the feasible region, as
\begin{equation} \label{eq_Py2_hat}
    \widehat{P}_y = \frac{1}{K} \sum_{k \in \mathcal{K}} \frac{1}{M} |\textbf{u}^H(\theta) \textbf{y}_k|^2,
\end{equation}
\begin{equation} \label{eq_sigma1_hat2}
    \widehat{\sigma}_1^2 = \frac{1}{NM-K} \Big( \sum_{k=1}^N \lVert \textbf{y}_k \rVert^2 - \sum_{k \in \mathcal{K}} \frac{1}{M} |\textbf{u}^H(\theta) \textbf{y}_k|^2 \Big).
\end{equation}
Substituting \eqref{eq_Py2_hat} and \eqref{eq_sigma1_hat2} into \eqref{eq_multi_antenna_numerator}, we obtain
\begin{equation} \label{eq_multi_antenna_numerator_2}
    \mathcal{L}_{\textrm{n}} = -NM \ln{\pi} -(NM-K)\ln{\widehat{\sigma}_1^2} -K\ln{\widehat{P}_y} -NM.
\end{equation}
Similarly, the denominator can be calculated by \eqref{eq_multi_antenna_likelihood_H0_final} as
\begin{equation} \label{eq_multi_antenna_denominator}
    \mathcal{L}_{\textrm{d}} = -NM \ln{\pi} - NM\ln{\sigma_0^2} - \sum_{k=1}^N \frac{\lVert \textbf{y}_k \rVert^2}{\sigma_0^2}.
\end{equation}
From the derivative test, the optimal value of $\sigma_0^2$ is obtained as
\begin{equation} \label{eq_multi_antenna_sigma_0_hat}
    \widehat{\sigma}_0^2 = \frac{1}{NM}\sum_{k=1}^N \lVert \textbf{y}_k \rVert^2.
\end{equation}
Plugging \eqref{eq_multi_antenna_sigma_0_hat} into \eqref{eq_multi_antenna_denominator}, the denominator simplifies to
\begin{equation} \label{eq_multi_antenna_denominator_2}
    \mathcal{L}_{\textrm{d}} = -NM \ln{\pi} - NM\ln{\widehat{\sigma}_0^2} - NM.
\end{equation}
Combining the calculations of the numerator and denominator in \eqref{eq_multi_antenna_numerator_2} and \eqref{eq_multi_antenna_denominator_2}, respectively, the GLRT $\mathcal{L}_4 = \mathcal{L}_{\textrm{n}} - \mathcal{L}_{\textrm{d}}$ can be written as
\begin{equation} \label{eq_multi_antenna_GLRT2_metric_2}
    \mathcal{L}_4 = \underset{\mathcal{K}, \theta}{\text{max} } \hspace{1mm} \Big(NM\ln{\widehat{\sigma}_0^2}-(NM-K)\ln{\widehat{\sigma}_1^2} -K\ln{\widehat{P}_y} \Big).
\end{equation}
Notably, for a given $K$, the expression inside the maximization is a monotonically increasing function of $\widehat{P}_y$ for $\widehat{P}_y \geq \widehat{\sigma}_1^2$, which is the region of interest. Consequently, for a given $K$, this GLRT selects the maximum $\widehat{P}_y$, similar to the previously proposed GLRT detectors.
As shown in Section~\ref{sec_multiantenna_extension}-\ref{sec_multi_antenna_GLRT_known_noise}, the maximum $\widehat{P}_y$ can be efficiently computed using the CUSUM operation, as detailed in \eqref{eq_multi_antenna_optimal_P}. Therefore, the multi-antenna GLRT with unknown noise power can be implemented in a manner analogous to the multi-antenna GLRT with known noise power presented in Algorithm~\ref{Algo_3}. The proposed multi-antenna GLRT with unknown noise power is summarized in Algorithm~\ref{Algo_4}.
%

\setlength{\textfloatsep}{2pt}
\begin{algorithm}[H]
{\fontsize{10pt}{10pt}\selectfont
\footnotesize
\caption{Multi-antenna GLRT with unknown noise power}\label{Algo_4}
\begin{algorithmic}[1]
\Require $\textbf{y}_k$ for $k=1,2,\dots,N$ in \eqref{eq_multi_antenna_signal_model}, Threshold: $\gamma_4$



    %
    
    \vspace{1mm}
    \State Calculate the CUSUM in \eqref{eq_multi_antenna_cusum}: $\textbf{C} \in \mathbb{R}^{M \times (N+1)}$
    

    \vspace{1mm}
    \State Calculate the average received power: $\widehat{\sigma}_0^2 = \frac{1}{NM}\sum_{k=1}^N \lVert \textbf{y}_k \rVert^2$
    
    \vspace{1mm}
    \For{$K \in \mathcal{S}_K = \{1,2,4,8,16,\dots,2^{\lfloor \log_2{N} \rfloor} \}$}
        \vspace{1mm}
        \State $P_K = \frac{1}{K} \lVert \textbf{C}_{(:,K+1:N+1)} - \textbf{C}_{(:,1:N-K+1)} \rVert_{\infty}$

        \vspace{1mm}
        \State $\widehat{\sigma}_1^2 = \frac{NM \widehat{\sigma}_0^2 - K P_K}{NM-K}$
        
        \vspace{1mm}
        \State $\mathcal{L}_4(K) = NM\ln{\widehat{\sigma}_0^2} -(NM-K)\ln{\widehat{\sigma}_1^2} -K\ln{P_K}$
        %
    
    \vspace{1mm}
    \EndFor 

    \vspace{1mm}
    \State Decision: $\mathcal{H}_1$ if $\max_{K \in \mathcal{S}_K} \{ \mathcal{L}_4(K) \} \geq \gamma_4$, and $\mathcal{H}_0$ otherwise
    
\vspace{1mm}    
\Output Decision: $\mathcal{H}_0 \textrm{ or } \mathcal{H}_1$
\end{algorithmic}}
\end{algorithm}

\vspace{-5mm}
\section{THEORETICAL ANALYSIS ON DETECTION} \label{sec_theoretical_analysis}
The theoretical analysis of the proposed GLRT detectors is intractable due to the search involved in their implementation. Instead, we analyze the baseline Conv-ED and the hypothetical UMP test to highlight the key properties an effective detector should exhibit.
%

\vspace{2mm}
\noindent {\textsc{\textbf{Lemma 1:}}} Let $T_G$ denote a Gaussian test statistic under both hypotheses. The probability of detection $P_{\textrm{d}} \triangleq \operatorname{Pr} \left(T_G \geq \gamma_G \mid \mathcal{H}_1 \right)$ for a constant false-alarm rate $P_{\textrm{fa}} \triangleq \operatorname{Pr} \left(T_G \geq \gamma_G \mid \mathcal{H}_0 \right)$ is given by
\begin{equation} \label{eq_Pd_final}
    P_{\textrm{d}} = Q \bigg( \frac{\sqrt{\operatorname{Var} \left\{ T_G \mid \mathcal{H}_0 \right\}}}{\sqrt{\operatorname{Var} \left\{ T_G \mid \mathcal{H}_1 \right\}}} Q^{-1}\left( P_{\textrm{fa}}\right) - D_G\bigg),
\end{equation}
where $Q(x) \triangleq \frac{1}{\sqrt{2\pi}} \int_{x}^{\infty} e^{-t^2/2} dt$ denotes the $Q$-function, and $D_G$ denotes the normalized distance, given by
\begin{equation}
    D_G \triangleq \frac{\mathbb{E} \left\{ T_G \mid \mathcal{H}_1 \right\} - \mathbb{E} \left\{ T_G \mid \mathcal{H}_0 \right\}}{\sqrt{\operatorname{Var} \left\{ T_G \mid \mathcal{H}_0 \right\}}}.
\end{equation}
This coefficient ($D_G$) can be interpreted as the effective detection SNR for the test $T_G$.
%

\vspace{2mm}
%
\noindent {\textit{\textbf{Proof:}}} From the well-known properties of Gaussian distributions, $P_{\textrm{fa}}$ and $P_{\textrm{d}}$ are given by
\begin{equation} \label{eq_Pfa}
    P_{\textrm{fa}} = Q \bigg( \frac{\gamma_G - \mathbb{E} \left\{ T_G \mid \mathcal{H}_0 \right\}}{\sqrt{\operatorname{Var} \left\{ T_G \mid \mathcal{H}_0 \right\}}}\bigg),
\end{equation}
\begin{equation} \label{eq_Pd}
    P_{\textrm{d}} = Q \bigg( \frac{\gamma_G - \mathbb{E} \left\{ T_G \mid \mathcal{H}_1 \right\}}{\sqrt{\operatorname{Var} \left\{ T_G \mid \mathcal{H}_1 \right\}}}\bigg).
\end{equation}
Consequently, the threshold corresponding to a given false-alarm rate $P_{\textrm{fa}}$ can be obtained from \eqref{eq_Pfa} as
\begin{equation} \label{eq_threshold}
    \gamma_G = \mathbb{E} \left\{ T_G \mid \mathcal{H}_0 \right\} + \sqrt{\operatorname{Var} \left\{ T_G \mid \mathcal{H}_0 \right\}} Q^{-1}\left( P_{\textrm{fa}}\right).
\end{equation}
Substituting \eqref{eq_threshold} into \eqref{eq_Pd} yields \eqref{eq_Pd_final}, which completes the proof.$\hfill \blacksquare$

\vspace{-3mm}
\subsection{Conventional Energy Detector (Conv-ED)} \label{sec_theoretical_ED}
The detection test for the Conv-ED is given by
\begin{equation} \label{eq_ED_test}
    T_{\textrm{ED}} = \frac{1}{MN \sigma^2} \sum_{k=1}^N \lVert \textbf{y}_k \rVert^2 \LRT{\mathcal{H}_1}{\mathcal{H}_0} \gamma_{\textrm{ED}}.
\end{equation}
Accordingly, we can obtain
\begin{equation} \label{eq_ED_mean_H0}
    \mathbb{E} \left\{ T_{\textrm{ED}} \mid \mathcal{H}_0 \right\} = 1,
\end{equation}
\begin{equation} \label{eq_ED_var_H0}
    \operatorname{Var} \left\{ T_{\textrm{ED}} \mid \mathcal{H}_0 \right\} = 1/{(MN)},
\end{equation}
\begin{equation} \label{eq_ED_mean_H1}
    \mathbb{E} \left\{ T_{\textrm{ED}} \mid \mathcal{H}_1 \right\} = 1 + {P_{\textrm{s}}}/{\sigma^2},
\end{equation}
\begin{equation} \label{eq_ED_var_H1}
    \operatorname{Var} \left\{ T_{\textrm{ED}} \mid \mathcal{H}_1 \right\} = \frac{1}{K} \Big(\frac{P_{\textrm{s}}}{\sigma^2}\Big)^2 + \frac{2}{NM} \Big(\frac{P_{\textrm{s}}}{\sigma^2}\Big) + \frac{1}{NM}.
\end{equation}
Importantly, since $N$ is large, the test statistic $T_{\textrm{ED}}$ in \eqref{eq_ED_test} can be approximated as Gaussian by the central limit theorem. Consequently, using Lemma \textcolor{blue}{1}, the probability of detection is given by
\begin{equation}
    P_{\textrm{d}} 
    = Q\Bigg( \frac{Q^{-1}(P_{\textrm{fa}}) - \sqrt{NM} \frac{P_{\textrm{s}}}{\sigma^2}}{\sqrt{ \frac{NM}{K} \left(\frac{P_{\textrm{s}}}{\sigma^2}\right)^2 + 2 \left(\frac{P_{\textrm{s}}}{\sigma^2}\right) + 1}} \Bigg).
\end{equation}
In wideband sensing, the SNR is typically low because most of the observed band contains only noise. Therefore, the terms $\frac{P_{\textrm{s}}}{\sigma^2}$ in the denominator become negligible. Then, $P_{\textrm{d}}$ can be approximated as
\begin{equation} \label{eq_Conv_ED_Pd}
    P_{\textrm{d}} \approx Q\Big( Q^{-1}(P_{\textrm{fa}}) - \sqrt{NM} \frac{P_{\textrm{s}}}{\sigma^2} \Big).
\end{equation}
Notably, the term $D_{\textrm{ED}} \triangleq \sqrt{NM} \frac{P_{\textrm{s}}}{\sigma^2}$ corresponds to the normalized distance for the Conv-ED.

\vspace{-5mm}
\subsection{Hypothetical Uniformly Most Powerful (UMP) Test} \label{sec_theoretical_UMP}
When the bandwidth $W$, the carrier frequency $f_c$, the signal power $P_{\textrm{s}}$, the noise power $\sigma^2$, and the AoA $\theta$ are all known, the UMP test can be expressed as
\begin{equation} \label{eq_UMP_test}
    T_{\textrm{UMP}} = \frac{1}{MK \sigma^2} \sum_{k \in \mathcal{K}} \lvert \textbf{u}^H(\theta) \textbf{y}_k \rvert^2 \LRT{\mathcal{H}_1}{\mathcal{H}_0} \gamma_{\textrm{UMP}}.
\end{equation}
Accordingly, we obtain
\begin{equation} \label{eq_UMP_mean_H0}
    \mathbb{E} \left\{ T_{\textrm{UMP}} \mid \mathcal{H}_0 \right\} = 1,
\end{equation}
\begin{equation} \label{eq_UMP_var_H0}
    \operatorname{Var} \left\{ T_{\textrm{UMP}} \mid \mathcal{H}_0 \right\} = \frac{1}{K},
\end{equation}
\begin{equation} \label{eq_UMP_mean_H1}
    \mathbb{E} \left\{ T_{\textrm{UMP}} \mid \mathcal{H}_1 \right\} = \frac{MN}{K} \frac{P_{\textrm{s}}}{\sigma^2} + 1,
\end{equation}
\begin{equation} \label{eq_UMP_var_H1}
    \operatorname{Var} \left\{ T_{\textrm{UMP}} \mid \mathcal{H}_1 \right\} = \frac{1}{K} \Big(\frac{MN}{K} \frac{P_{\textrm{s}}}{\sigma^2} + 1 \Big)^2.
\end{equation}
Importantly, for large $K$ values, the test statistic $T_{\textrm{UMP}}$ in \eqref{eq_UMP_test} can be approximated as Gaussian according to the central limit theorem. Consequently, from Lemma \textcolor{blue}{1}, the probability of detection is given by
\begin{equation}
    P_{\textrm{d}}
    = Q\bigg( \frac{Q^{-1}(P_{\textrm{fa}}) - \frac{MN}{\sqrt{K}} \frac{P_{\textrm{s}}}{\sigma^2}}{\frac{MN}{K} \frac{P_{\textrm{s}}}{\sigma^2} + 1} \bigg).
\end{equation}
At low SNR, the probability of detection $P_{\textrm{d}}$ can be approximated as
\begin{equation} \label{eq_UMP_Pd}
    P_{\textrm{d}} \approx Q\bigg( Q^{-1}(P_{\textrm{fa}}) - \frac{MN}{\sqrt{K}} \frac{P_{\textrm{s}}}{\sigma^2} \bigg).
\end{equation}
Notably, the term $D_{\textrm{UMP}} \triangleq \frac{MN}{\sqrt{K}} \frac{P_{\textrm{s}}}{\sigma^2}$ in \eqref{eq_UMP_Pd} corresponds to the normalized distance for the UMP test.

Importantly, in the single-antenna setting, if the signal bandwidth matches the observed band ($K = N$), the Conv-ED test statistic in \eqref{eq_ED_test} coincides with the hypothetical UMP in \eqref{eq_UMP_test}, making the Conv-ED optimal \cite{ED_14}. However, in blind sensing, the observed band is typically much larger than the signal band. Consequently, the Conv-ED deviates from optimality and its performance degrades, motivating the need for an efficient detector that can accurately identify the signal band.

\vspace{-5mm}
\subsection{Impact of System Parameters on Detection} \label{sec_phys_params_impact}
This section examines the impact of key system parameters on the detection performance of both the Conv-ED and the hypothetical UMP test. From \eqref{eq_Conv_ED_Pd} and \eqref{eq_UMP_Pd}, for a fixed detection probability $P_{\textrm{d}}$ at constant false-alarm rate $P_{\textrm{fa}}$, the normalized distances $D_{\textrm{ED}}$ and $D_{\textrm{UMP}}$ must remain constant.
%
%
Below, we analyze the impact of system parameters on the normalized distances. 
%
%
\begin{itemize}
    \item \textbf{Effect of SNR:} The normalized distances scale as
    \begin{equation}
        D_{\textrm{ED}} \propto \textrm{SNR} \quad , \quad 
        D_{\textrm{UMP}} \propto \textrm{SNR} .
    \end{equation}
    Hence, increasing the SNR improves the detection performance of both the Conv-ED and the UMP.
    
    \item \textbf{Effect of bandwidth:} The bandwidth $W$ of the target signal affects the interval length $K = \lfloor WT \rfloor$. Consequently, the normalized distances scale as
    \begin{equation}
        D_{\textrm{ED}} \propto 1 \quad , \quad 
        D_{\textrm{UMP}} \propto \tfrac{1}{\sqrt{W}} .
    \end{equation}
    Therefore, as the bandwidth of the target signal decreases, the detection performance of the UMP improves, whereas the Conv-ED remains insensitive.

    \item \textbf{Effect of observation duration:} The observation duration $T$ affects both the number of samples $N = \lfloor f_s T \rfloor$ and the interval length $K = \lfloor WT \rfloor$. Consequently, the normalized distances scale as
    \begin{equation}
        D_{\textrm{ED}} \propto \sqrt{T} \quad , \quad 
        D_{\textrm{UMP}} \propto \sqrt{T} .
    \end{equation}
    Hence, increasing the observation duration improves the performance of both the Conv-ED and the UMP.

    \item \textbf{Effect of sampling rate:} The sampling rate $f_s$ affects both the number of samples $N = \lfloor f_s T \rfloor$ and the noise power $\sigma^2 = N_0 f_s$. Consequently, the normalized distances scale as
    \begin{equation}
        D_{\textrm{ED}} \propto \tfrac{1}{\sqrt{f_s}} \quad , \quad 
        D_{\textrm{UMP}} \propto 1 .
    \end{equation}
    Hence, increasing the sampling rate degrades the detection performance of the Conv-ED, while the UMP remains largely insensitive.

    \item \textbf{Effect of number of antennas:} The normalized distances scale as
    \begin{equation}
        D_{\textrm{ED}} \propto \sqrt{M} \quad , \quad 
        D_{\textrm{UMP}} \propto M .
    \end{equation}
    Hence, increasing the number of antennas improves the detection performance of both the Conv-ED and the UMP, with the UMP benefiting more efficiently from the increase in $M$.
\end{itemize}
Let $P_{\textrm{req}}$ and $T_{\textrm{req}}$ denote the required signal power $(P_{\textrm{s}})$ and observation duration $(T)$, respectively, to maintain a given detection probability. Based on the analysis above, the effect of system parameters on these required values is summarized in Table \ref{table_impact_of_params}.

\vspace{1mm}
\begin{table}[ht]
\caption{Impact of System Parameters}
\label{table_impact_of_params}
\centering
\setlength\tabcolsep{4pt}
\renewcommand{\arraystretch}{1.7}
\resizebox{1\columnwidth}{!}{
\begin{tabular}{|c|c|c|}
\hline
\multicolumn{1}{|c|}{System parameters} & \multicolumn{1}{c|}{Conv-ED}                            & UMP                                                   \\ \hline
SNR $\frac{P_{\textrm{s}}}{\sigma^2}$                            & $T_{\textrm{req}} \propto \frac{1}{{\textrm{SNR}}^2}$                                                  & $T_{\textrm{req}} \propto \frac{1}{{\textrm{SNR}}^2}$                                                \\ \hline
Bandwidth $W$                               & $P_{\textrm{req}} \propto 1$ ; $T_{\textrm{req}} \propto 1$& $P_{\textrm{req}} \propto \sqrt{W}$ ; $T_{\textrm{req}} \propto W$\\ \hline
Obs. duration $T$                    & $P_{\textrm{req}} \propto \frac{1}{\sqrt{T}}$                                                  & $P_{\textrm{req}} \propto \frac{1}{\sqrt{T}}$                                                  \\ \hline
Sampling rate $f_s$                          & $P_{\textrm{req}} \propto \sqrt{f_s}$ ; $T_{\textrm{req}} \propto f_s$& $P_{\textrm{req}} \propto 1$ ;  $T_{\textrm{req}} \propto 1$\\ \hline
No. of antennas $M$                      & $P_{\textrm{req}} \propto \frac{1}{\sqrt{M}}$ ; $T_{\textrm{req}} \propto \frac{1}{M}$& $P_{\textrm{req}} \propto \frac{1}{M}$ ; $T_{\textrm{req}} \propto \frac{1}{M^2}$\\ \hline
\end{tabular}
}
\end{table}

Notably, the SNR gap between the Conv-ED and the UMP is given by $\frac{D_{\textrm{UMP}}}{D_{\textrm{ED}}}=\sqrt{\frac{MN}{K}}$, indicating that Conv-ED becomes ineffective in wideband sensing scenarios ($N \gg K$) and in high-dimensional spatial regimes ($M \gg 1$).

\vspace{-3mm}
\section{COMPLEXITY ANALYSIS} \label{sec_complexity}
The proposed multi-antenna GLRT detectors first compute the CUSUM matrix with a complexity of $\mathcal{O}(NM \log M)$. They then search over $|\mathcal{S}_K| = \lfloor \log_2 N \rfloor$ candidate interval lengths $K$. For each candidate, computing the CUSUM differences requires $\mathcal{O}(NM)$ operations, resulting in an overall search complexity of $\mathcal{O}(NM \log N)$. Hence, the total complexity of the multi-antenna GLRT detectors is $\mathcal{O}(NM \log NM)$. The single-antenna GLRT detectors correspond to $M=1$, yielding a complexity of $\mathcal{O}(N \log N)$.

In the simulation results, we consider two well-known conventional detection methods for comparison. The first is the Conv-ED, and the second is the conventional peak detector (Conv-PD). The computational complexity of the Conv-ED is $\mathcal{O}(MN)$.
The Conv-PD divides the received signal at each antenna into segments of length $N_{\textrm{fft}}$, computes the FFT for each segment, and accumulates the squared magnitudes across segments and antennas, corresponding to the Bartlett/Welch PSD estimate. The maximum PSD value is then compared to a threshold. With $\lfloor MN / N_{\textrm{fft}} \rfloor$ segments, the overall complexity is $\mathcal{O}(MN \log N_{\textrm{fft}}) \approx \mathcal{O}(MN \log N)$.
The computational complexity orders are summarized in Table~\ref{table_complexity}.

\begin{table}[ht]
\caption{Computational Complexity Orders}
\label{table_complexity}
\centering
\setlength\tabcolsep{4pt}
\renewcommand{\arraystretch}{1.2}
\resizebox{0.9\columnwidth}{!}{
\begin{tabular}{|c|c|c|}
\hline
 & \textbf{Method} & \textbf{Complexity} \\ \hline \hline

\multirow{4}{*}[-1.7ex]{\centering\rotatebox[origin=c]{90}{Single-antenna}}
& Conv-ED & $O(N)$ \\ \cline{2-3}
& Conv-PD & $O(N\log N)$ \\ \cline{2-3}
& \makecell{Algo.~\ref{Algo_1}: GLRT with known\\ noise power (proposed)} 
& $O(N\log N)$ \\ \cline{2-3}
& \makecell{Algo.~\ref{Algo_2}: GLRT with unknown\\ noise power (proposed)} 
& $O(N\log N)$ \\ \hline \hline

\multirow{4}{*}[-1.9ex]{\centering\rotatebox[origin=c]{90}{Multi-antenna}}
& Conv-ED & $O(MN)$ \\ \cline{2-3}
& Conv-PD & $O(MN\log N)$ \\ \cline{2-3}
& \makecell{Algo.~\ref{Algo_3}: Multi-antenna GLRT\\ with known noise power (proposed)} 
& $O(MN\log(MN))$ \\ \cline{2-3}
& \makecell{Algo.~\ref{Algo_4}: Multi-antenna GLRT\\ with unknown noise power (proposed)} 
& $O(MN\log(MN))$ \\ \hline 

\end{tabular}
}
\end{table}

\vspace{-4mm}
\section{SIMULATION RESULTS} \label{sec_sim_results}
\begin{figure}[ht]
    \centering
    \includegraphics[width=0.999\columnwidth]{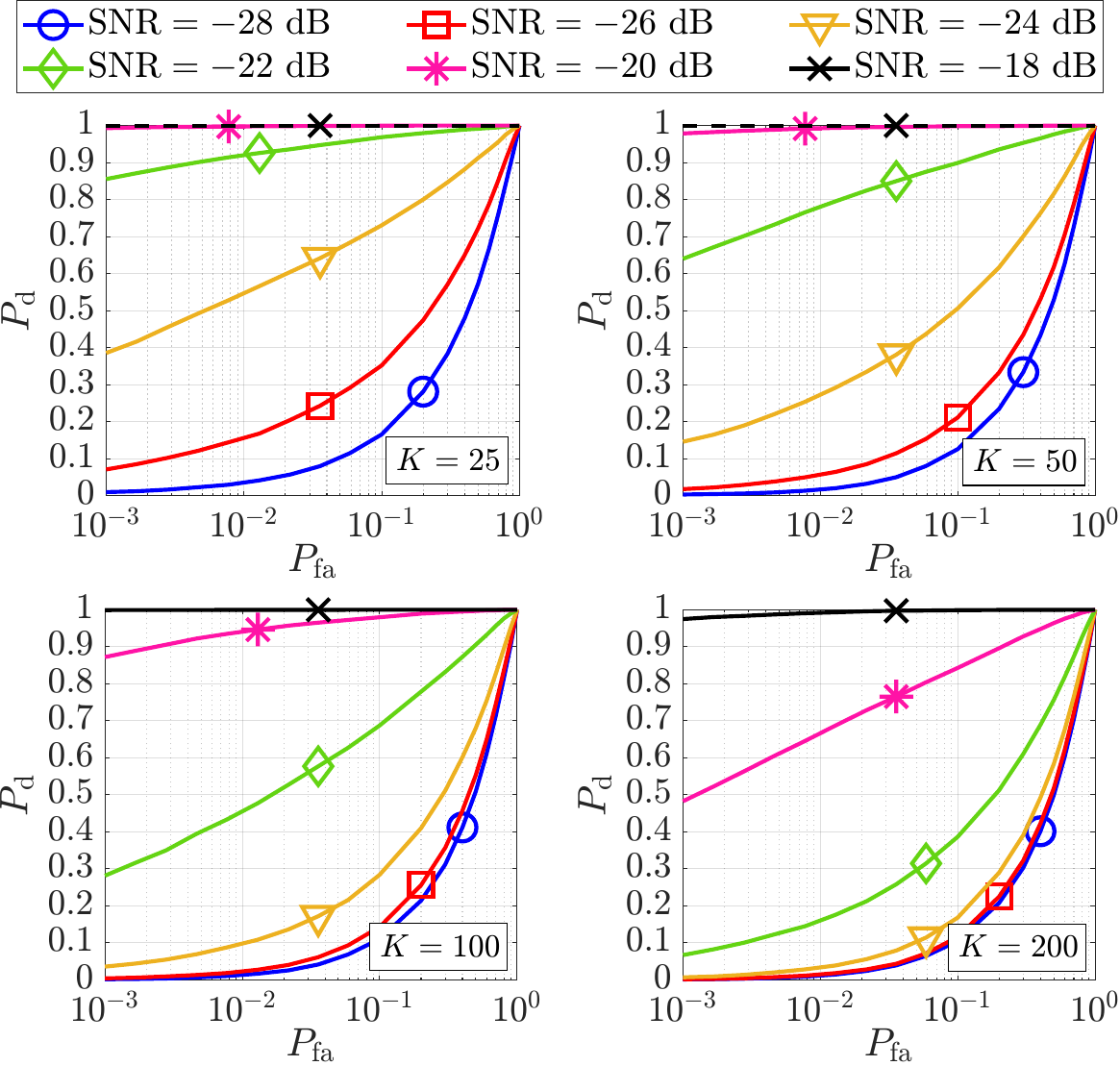}
    \caption{ROC of the proposed GLRT with unknown noise power}
    \label{fig_ROC}
\end{figure}

\begin{figure}[ht]
    \centering
    \includegraphics[width=0.999\columnwidth]{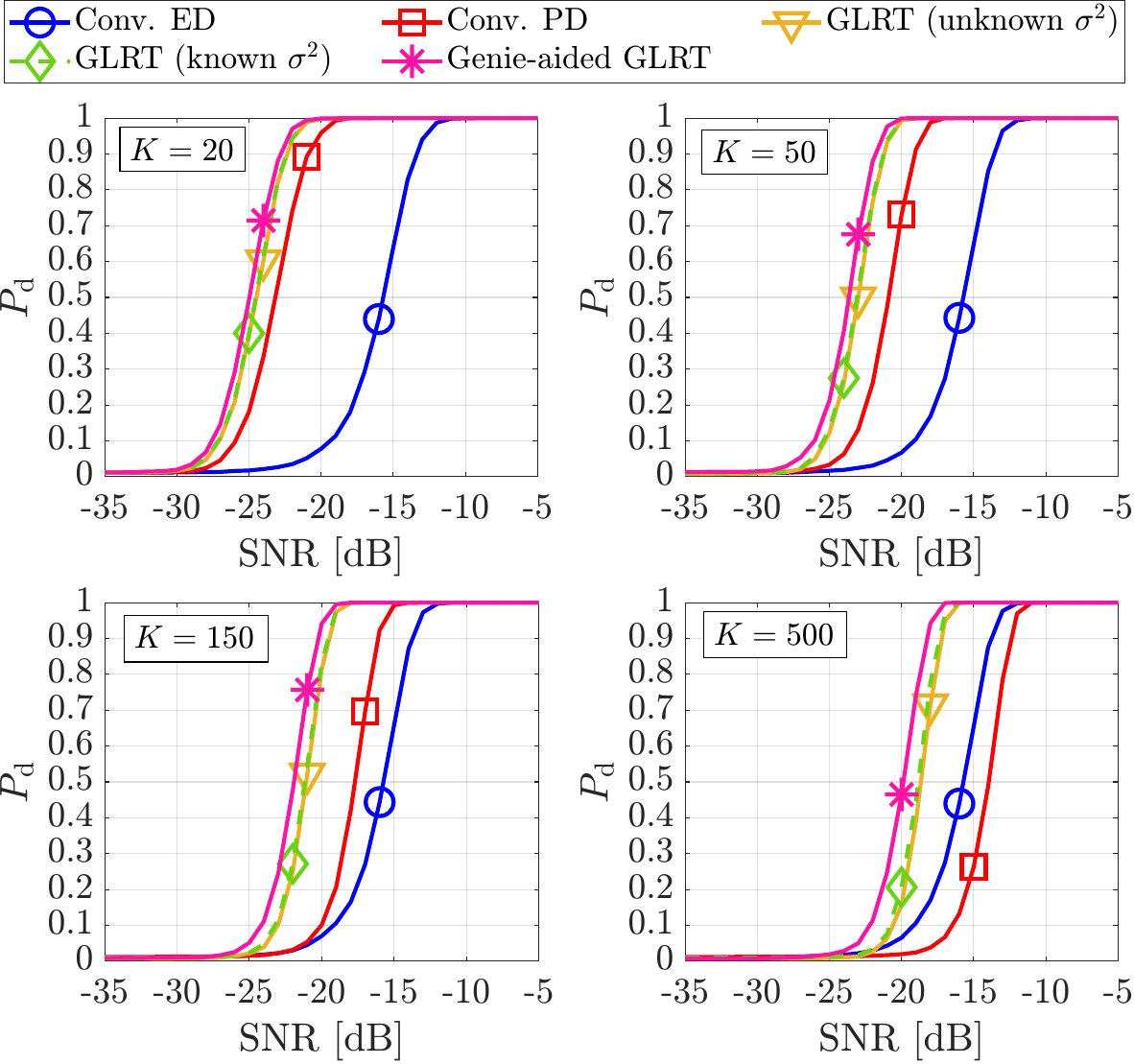}
    \caption{Probability of detection $P_{\textrm{d}}$ vs. SNR for various detector designs (The sampling rate $f_s$ and the observation duration $T$ are kept fixed)}
    \label{fig_Pd_vs_SNR}
    \vspace{3mm}
\end{figure}

\begin{figure}[ht]
    \centering
    
    \begin{subfigure}{0.999\columnwidth}
        \centering
        \includegraphics[width=1\columnwidth]{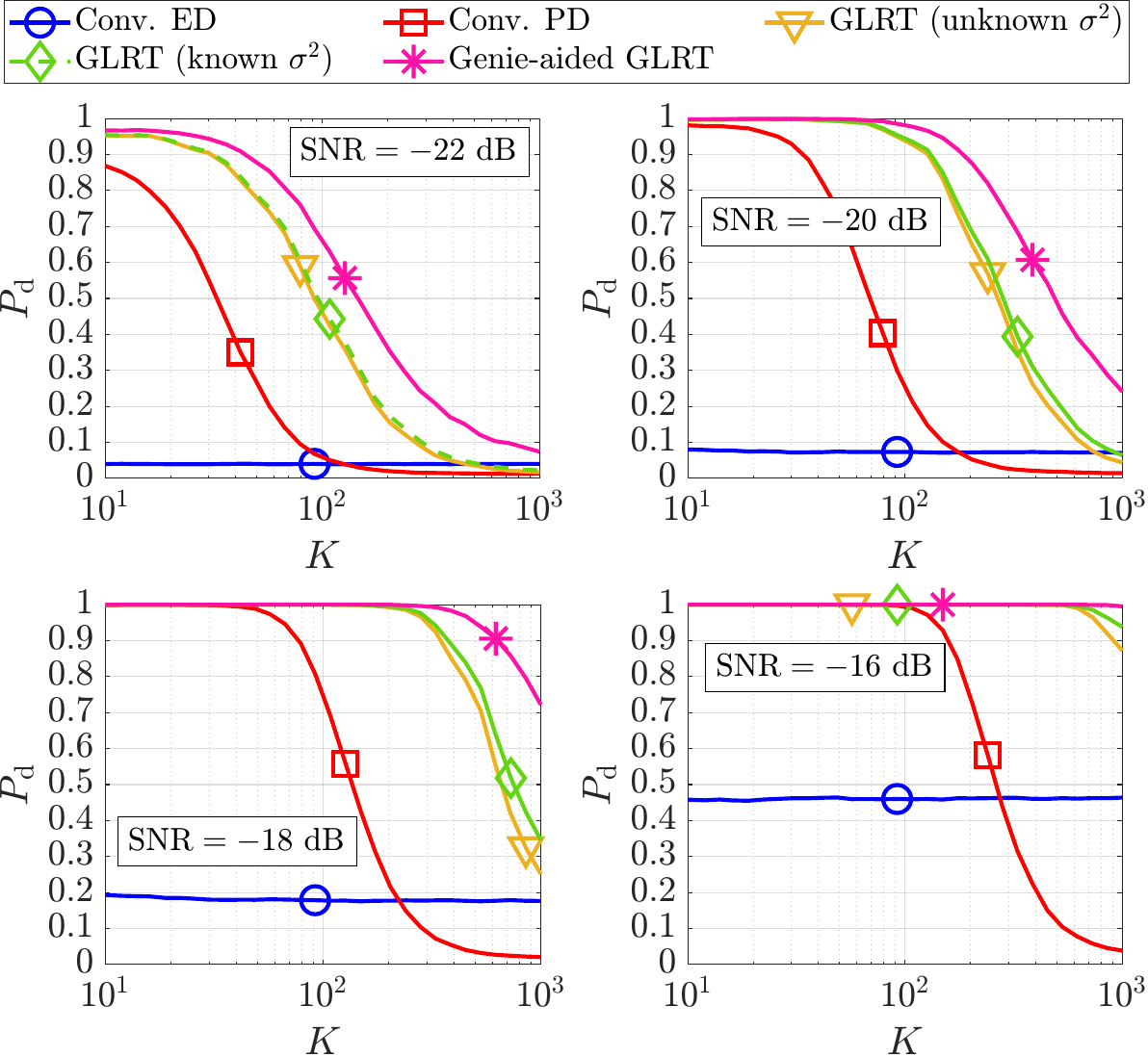}
        \caption{$P_{\textrm{d}}$ vs. $K$}
        \label{fig_Pd_vs_K}
    \end{subfigure}
    
    \vspace{0.5mm}

    \begin{subfigure}{0.999\columnwidth}
        \centering
        \includegraphics[width=0.5\columnwidth]{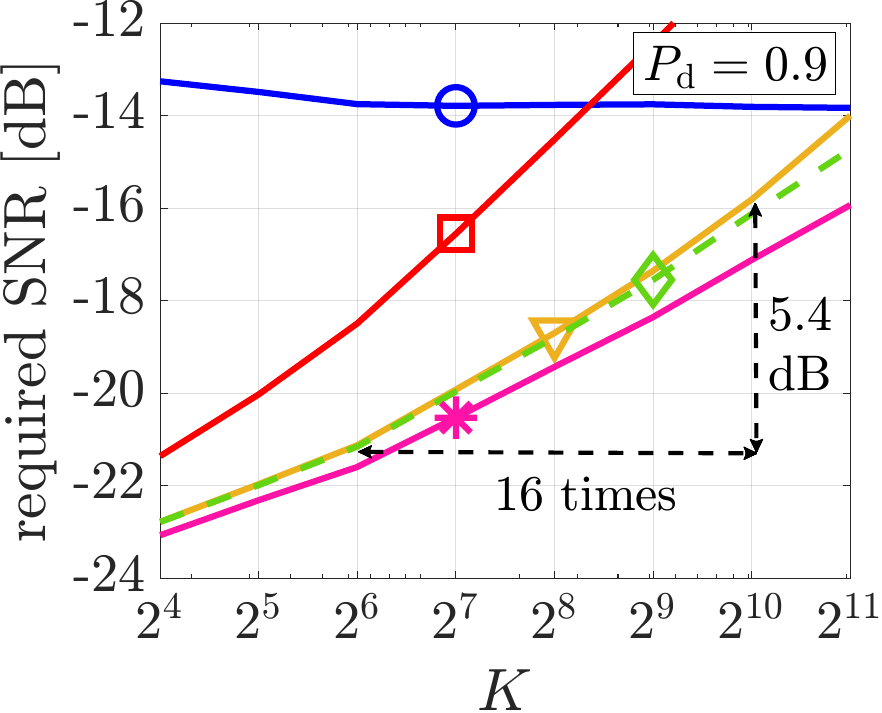}
        \caption{Required SNR vs. $K$}
        \label{fig_required_SNR_vs_K}
    \end{subfigure}
    
    \caption{Detection performance with respect to the bandwidth $W$; different $K$ values correspond to different bandwidth values (The sampling rate $f_s$ and the observation duration $T$ are kept fixed)}
    \label{fig_W}
    \vspace{3mm}
\end{figure}

\begin{figure}[ht]
    \centering
    \includegraphics[width=0.999\columnwidth]{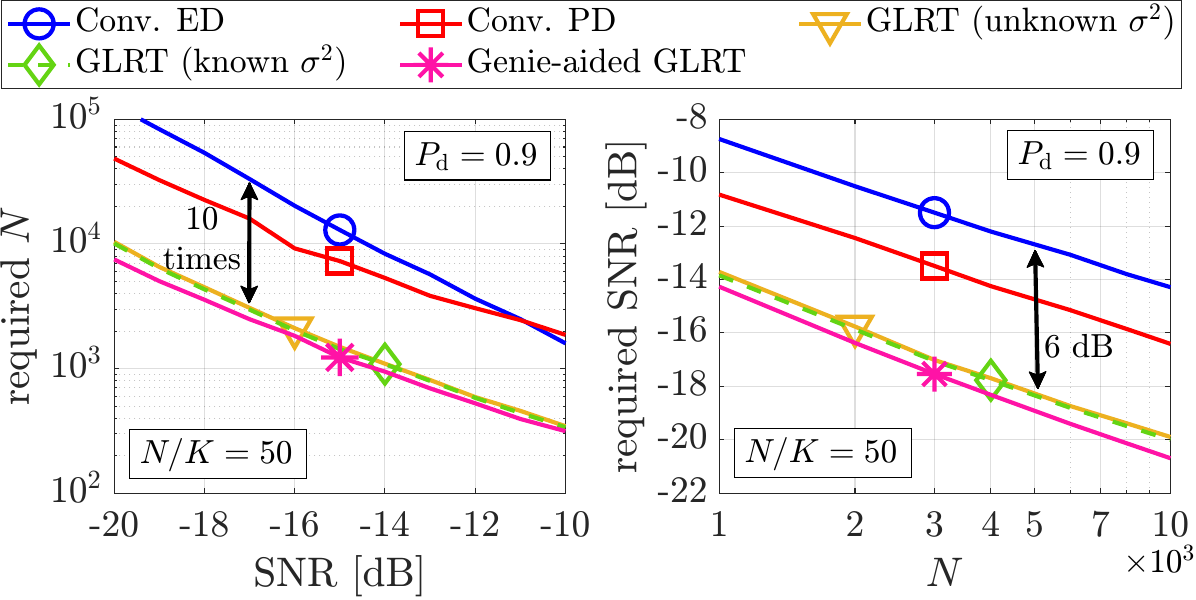}
    \caption{Detection performance with respect to the observation duration $T$; different $N$ values correspond to different observation durations (The sampling rate $f_s$ is kept fixed)}
    \label{fig_SNR_vs_required_N}
    \vspace{3mm}
\end{figure}

\begin{figure}[ht]
    \centering
    
    \begin{subfigure}{0.999\columnwidth}
        \centering
        \includegraphics[width=1\linewidth]{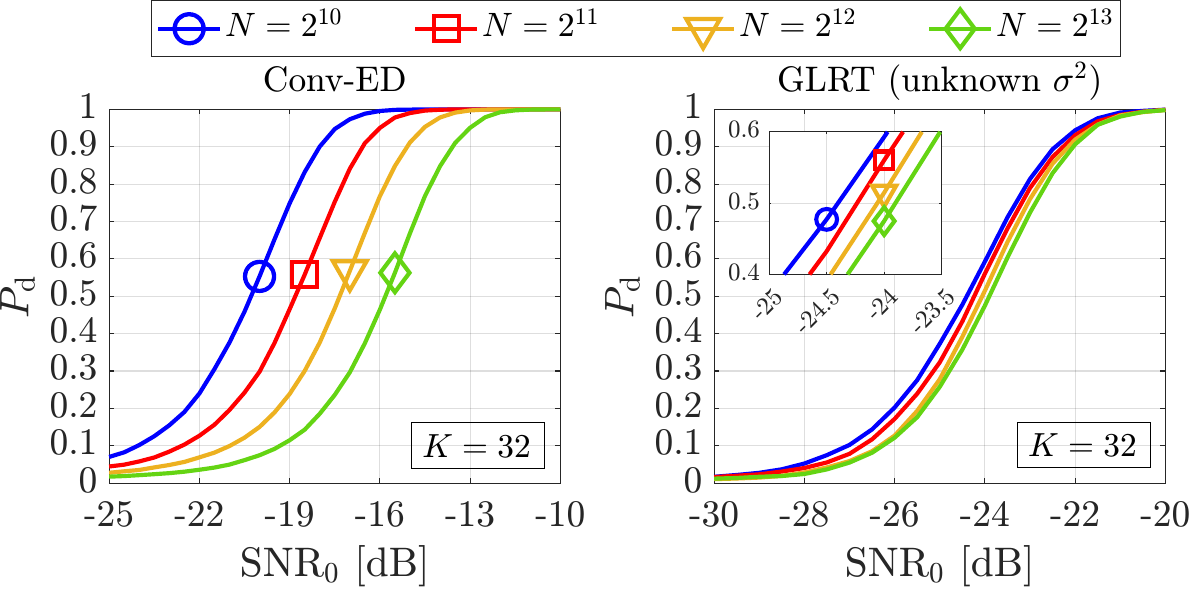}
        \caption{$P_{\textrm{d}}$ vs. $\textrm{SNR}_0$ for the Conv-ED and the proposed GLRT in Algo. \ref{Algo_2}}
        \label{fig_Pd_vs_SNR0_wrt_N}
    \end{subfigure}
    
    \vspace{0.5mm}

    \begin{subfigure}{0.999\columnwidth}
        \centering
        \includegraphics[width=1\linewidth]{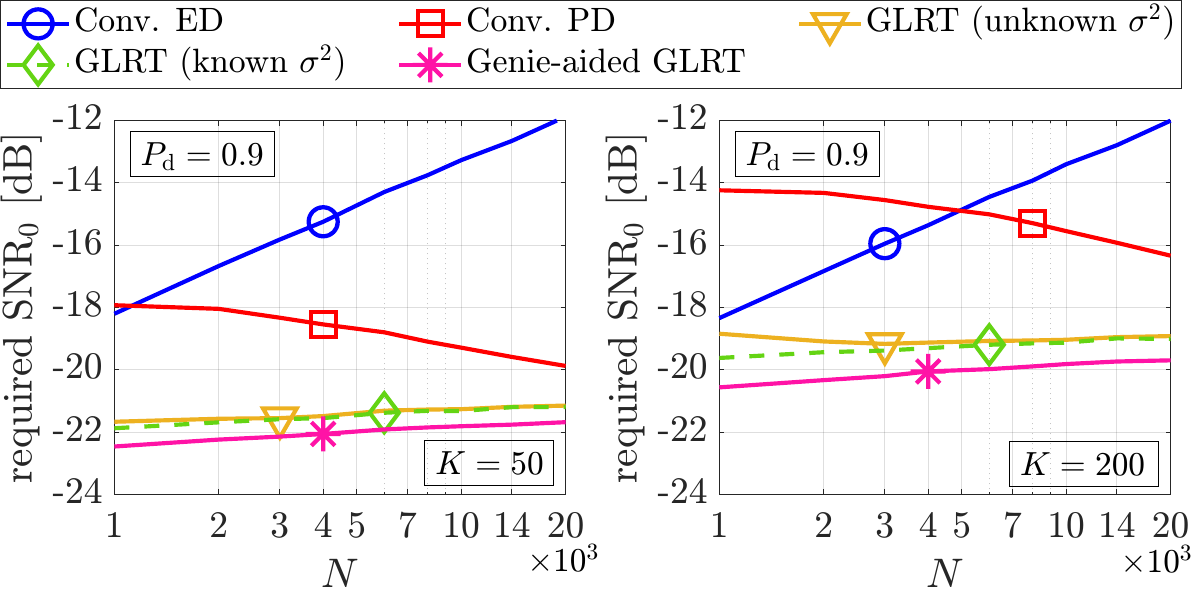}
        \caption{Required $\textrm{SNR}_0$ to achieve $P_{\textrm{d}}=0.9$ vs. $N$}
        \label{fig_required_SNR0_vs_N}
    \end{subfigure}
    
    \caption{Detection performance with respect to the sampling rate $f_s$; different $N$ values correspond to different sampling rates (The observation duration $T$ is kept fixed)}
    \label{fig_fs}
    \vspace{2mm}
\end{figure}

\begin{figure}[ht]
    \centering
    
    \begin{subfigure}{0.999\columnwidth}
        \centering
        \includegraphics[width=1\columnwidth]{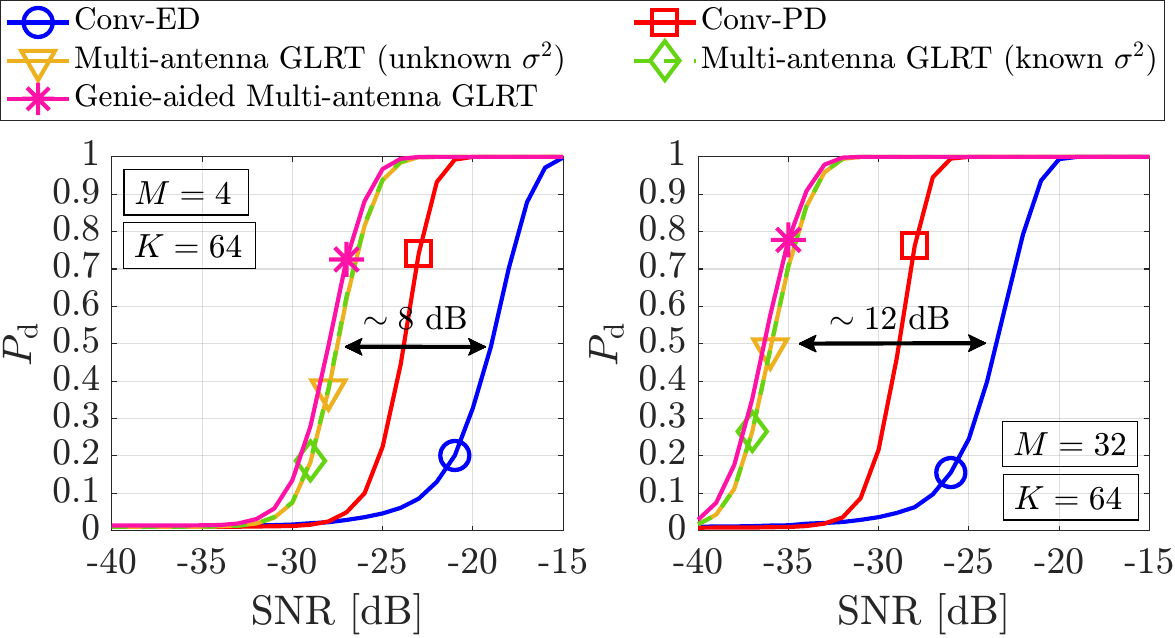}
        \caption{$P_{\textrm{d}}$ vs. SNR}
        \label{fig_Pd_vs_SNR_multiantenna_all_methods_comparison}
    \end{subfigure}
    
    \vspace{0.5mm}

    \begin{subfigure}{0.999\columnwidth}
        \centering
        \includegraphics[width=1\columnwidth]{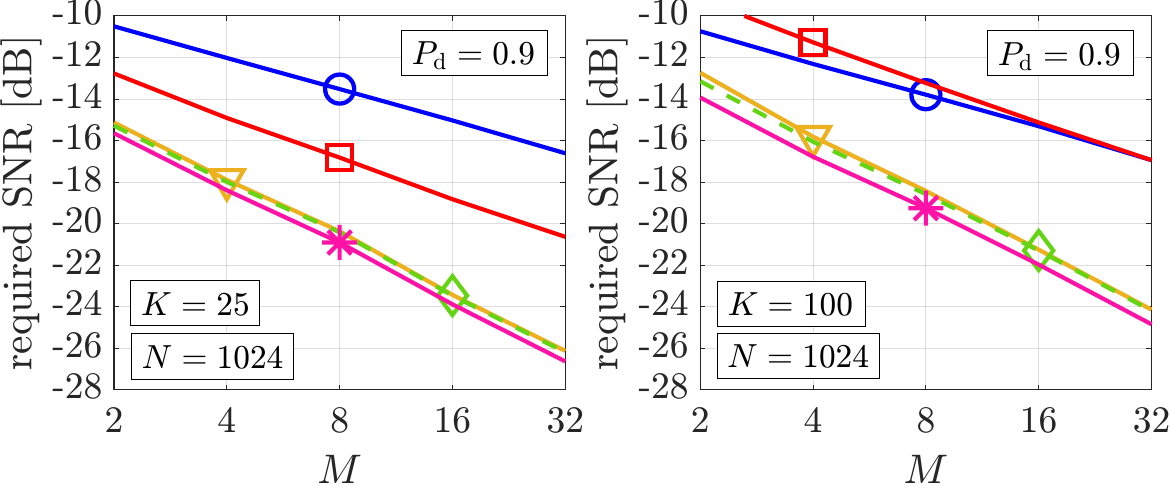}
        \caption{Required SNR to achieve $P_{\textrm{d}}=0.9$ vs. $M$}
        \label{fig_required_SNR_vs_M}
    \end{subfigure}
    
    \caption{Detection performance w.r.t. the number of antennas $M$}
    \label{fig_multi_antenna}
\end{figure}

In the simulations, we evaluate the probability of detection ($P_{\textrm{d}}$) of the considered detectors. For benchmarking purposes, we also include a genie-aided GLRT detector with perfect knowledge of the noise power $\sigma^2$, the signal power $P_{\textrm{s}}$, and the interval length $K$ (or equivalently the bandwidth $W$), while the starting bin index $k_0$ (or equivalently the carrier frequency $f_c$) remains unknown. In the simulations, Figs.~\ref{fig_ROC}, \ref{fig_Pd_vs_SNR}, \ref{fig_W}, \ref{fig_SNR_vs_required_N}, and \ref{fig_fs} consider a single-antenna setting, whereas Fig.~\ref{fig_multi_antenna} considers a multi-antenna setting. The detection thresholds for the considered detectors are determined numerically via Monte Carlo simulations to ensure a constant false-alarm rate $P_{\textrm{fa}}$. The number of received samples is set to $N=2^{13}=8192$, and the false-alarm rate is fixed at $P_{\textrm{fa}}=10^{-2}$. For a given interval length $K$, the starting bin index $k_0$ is generated at random from $\mathcal{U}[1, N-K+1]$. In the multi-antenna setting, the AoA is generated as $\widetilde{\theta} \sim \mathcal{U}[-45^{\circ}, 45^{\circ}]$. The FFT size for the Conv-PD is set to $N_{\textrm{fft}}=1024$. Unless otherwise specified, these parameter values are used throughout the simulations.



Fig.~\ref{fig_ROC} illustrates the receiver operating characteristic (ROC) of the proposed GLRT detector with unknown noise power in Algorithm~\ref{Algo_2}. The results show that, for a fixed interval length $K$, the probability of detection $P_{\textrm{d}}$ improves as the SNR increases, as expected.
The results also reveal that as $K$ increases while the SNR remains fixed, the ROC curves exhibit a systematic degradation. This occurs because the signal power is spread across a larger number of spectral bins, effectively reducing the SNR per bin and making the signal appear more noise-like.
This behavior is consistent with the fundamental principles of DSSS systems, which intentionally spread the signal over a wider bandwidth to reduce the PSD, thereby achieving a low probability of intercept (LPI).

\vspace{3mm}

Fig.~\ref{fig_Pd_vs_SNR} compares the detection performance of the considered detectors as a function of SNR for various interval lengths $K$ (equivalently, bandwidths $W$).
The results demonstrate that the proposed GLRT with unknown noise power provides a substantial SNR gain compared to the Conv-ED, particularly at small bandwidth values. Notably, while the Conv-PD performs well at small bandwidths, where the signal spectrum is relatively peaky, its performance degrades as $K$ increases. This degradation occurs because, as the signal bandwidth increases, the signal becomes more noise-like, resulting in the Conv-PD being increasingly susceptible to noise-only outliers that exceed the signal peak.
Furthermore, the negligible performance gap between the GLRT with known noise power and the GLRT with unknown noise power indicates that the proposed detector estimates the noise power effectively. Most remarkably, the performance of the proposed GLRT closely coincides with that of the genie-aided GLRT, thereby highlighting the robustness and effectiveness of the proposed detection framework.

\vspace{2mm}

Fig.~\ref{fig_W} illustrates the effect of the target signal bandwidth $W$ on the detection performance. Variations in $W$ are reflected through the interval length $K = \lfloor TW \rfloor$. Fig.~\ref{fig_Pd_vs_K} shows that the detection performance of the Conv-ED is largely insensitive to the target signal bandwidth. This observation is consistent with theory, since the Conv-ED performance depends primarily on the total signal power and not on the number of DFT bins over which the power is spread, as shown in Section~\ref{sec_theoretical_analysis} and Table~\ref{table_impact_of_params}.
Another important observation is that the performance of the GLRT-based detectors approaches that of the Conv-ED as $K$ increases. This is expected because when the bandwidth of the target signal coincides with the observed band, the Conv-ED becomes optimal \cite{ED_14}, as shown in Section~\ref{sec_theoretical_analysis}.
%
%
Fig.~\ref{fig_required_SNR_vs_K} shows the required SNR to achieve $P_{\textrm{d}} = 0.9$ as a function of the interval length $K$. As indicated in Table~\ref{table_impact_of_params}, the required SNR for the hypothetical UMP scales as $\sqrt{W}$, or equivalently $\sqrt{K}$. This implies that the required SNR should increase by approximately 6 dB when the interval length $K$ increases by a factor of 16. In the simulation results, the required SNR increases by 5.4 dB for the same change in $K$, demonstrating that the proposed GLRT detectors follow a trend similar to that of the hypothetical UMP test. 
Overall, Fig.~\ref{fig_W} shows that the proposed GLRT detectors become increasingly important in wideband spectrum sensing scenarios, where the observed band is substantially larger than the target signal bandwidth, which is typical in blind sensing.

\vspace{2mm}

Fig.~\ref{fig_SNR_vs_required_N} illustrates the effect of the observation duration $T$ on detection performance. Variations in $T$ are reflected through both the number of samples, $N = \lfloor T f_s \rfloor$, and the interval length, $K = \lfloor T W \rfloor$. Thus, both $N$ and $K$ are increased proportionally, keeping their ratio $N/K$ constant.
%
%
According to the theoretical analysis in Section~\ref{sec_theoretical_analysis}, the required $N$ should decrease by a factor of 100 when the SNR is increased by a factor of 10 (10 dB) for both the Conv-ED and the hypothetical UMP. The results for the Conv-ED closely follow the theoretical prediction, while the proposed GLRT also follows the theoretical trend of the UMP fairly closely.
Similarly, the theoretical analysis in Section~\ref{sec_theoretical_analysis} suggests that the required SNR should decrease by a factor of $\sqrt{10}$ (5 dB) when $N$ (or equivalently $T$) increases by a factor of 10, for both the Conv-ED and the hypothetical UMP. The simulation results follow this trend closely, with the required SNR decreasing by 5–6 dB for the Conv-ED and the proposed GLRT when $N$ increases 10 times.

\vspace{2mm}

Fig.~\ref{fig_fs} illustrates the effect of the sampling rate $f_s$ on detection performance. Variations in $f_s$ are reflected through the number of samples $N = \lfloor T f_s \rfloor$. Notably, changes in $f_s$ do not affect the interval length $K \triangleq \lfloor T W \rfloor$, since $K$ depends only on the target signal bandwidth $W$ and the observation duration $T$.
In the previous results, the sampling rate was fixed, making it appropriate to fix the SNR. In contrast, when comparing the impact of $f_s$, a fair comparison requires fixing the ratio $\frac{P_{\textrm{s}}}{N_0}$ across different values of $N$, as discussed in Section~\ref{sec_system_model}-\ref{sec_SNR_def}. Since $\frac{P_{\textrm{s}}}{N_0} = f_s \times \textrm{SNR}$, this implies that $f_s \times \textrm{SNR}$, or equivalently $N \times \textrm{SNR}$, must remain constant. To this end, we define $\textrm{SNR}_0 \triangleq \textrm{SNR} \times \frac{N}{8192}$ and keep $\textrm{SNR}_0$ fixed for different values of $N$. The constant 8192 serves solely as a normalization factor, ensuring that $\textrm{SNR} = \textrm{SNR}_0$ under the default setting $N = 2^{13} = 8192$.
From an information-theoretic perspective, increasing the sampling rate does not degrade performance, as more data samples are available. However, since the number of in-band DFT bins $K$ remains unchanged, the additional band contains only noise, offering no performance gain either. Hence, for an efficient detector, detection performance is expected to remain approximately invariant to the sampling rate, consistent with the theoretical UMP analysis in Table~\ref{table_impact_of_params}.
Fig.~\ref{fig_fs} indicates that increasing $N$ has only a marginal impact on the detection performance of the proposed GLRT detector. In fact, slightly lower detection probabilities are observed for larger values of $N$. This behavior is expected, since increasing $N$ enlarges the search space in the bandwidth and starting-bin-index ($k_0$) search stage, which in turn results in a modest degradation in detection performance. Overall, this observation is consistent with the earlier information-theoretic discussion and closely aligns with the theoretical trend of the UMP.
On the other hand, the performance of the Conv-ED is significantly affected by the sampling rate. This is because, for a fixed target signal, observing a wider band causes the Conv-ED to include additional noise in the energy calculation, which consequently degrades its performance, as further evident in Table~\ref{table_impact_of_params}.
This figure addresses the fundamental question of whether detection performance can be improved by increasing the sampling rate $f_s$. The results show that increasing $f_s$ has almost no effect on the detection performance of the proposed GLRT detector, similar to the hypothetical UMP test, whereas Conv-ED becomes increasingly ineffective. This is a significant distinction because, in practice, the observed band, or equivalently the sampling rate, is often chosen as large as possible in blind sensing scenarios to avoid missing the signal.

\vspace{2mm}



Fig.~\ref{fig_multi_antenna} illustrates the effect of the number of antennas $M$ on detection performance. Fig.~\ref{fig_Pd_vs_SNR_multiantenna_all_methods_comparison} shows that the performance gap between the Conv-ED and the proposed GLRT detectors increases as the number of antennas increases.
This widening margin arises because the proposed GLRT effectively exploits the full coherent array gain of order $O(M)$, whereas the conventional ED scales only as $O(\sqrt{M})$ due to non-coherent energy accumulation across antennas, as derived in Section~\ref{sec_theoretical_analysis}.
%
%
The theoretical analysis in Section~\ref{sec_theoretical_analysis} implies that the required SNR to achieve a fixed detection probability decreases by $10\log_{10}\left(\sqrt{4}\right) \approx 3$ dB for the Conv-ED and $10\log_{10}(4) \approx 6$ dB for the UMP when $M$ increases by a factor of four. Fig.~\ref{fig_required_SNR_vs_M} shows very similar results: the required SNR decreases by approximately 3 dB for the Conv-ED and 6 dB for the proposed GLRT detectors when $M$ increases from 8 to 32.
These results closely match the theoretical analysis, indicating that the proposed GLRT detectors follow the same scaling trend as the hypothetical UMP. Moreover, the negligible gap between the proposed GLRT with unknown noise power and the genie-aided GLRT further highlights the effectiveness of the proposed detector, even in high-dimensional spatial regimes.

\vspace{2mm}

\section{CONCLUSION} \label{sec_conclusion}
This study introduces efficient GLRT detectors for the blind detection of OFDM signals with unknown parameters in wideband spectrum sensing.
The results show that the performance of the proposed GLRT detectors follows a similar trend to the hypothetical UMP test. This indicates that the effective detection SNR of the proposed GLRT detectors scales as $\mathcal{O}\big(\frac{MN}{\sqrt{K}} \frac{P_{\textrm{s}}}{\sigma^2}\big)$, whereas the effective detection SNR of the Conv-ED scales as $\mathcal{O}\big(\sqrt{NM}\frac{P_{\textrm{s}}}{\sigma^2}\big)$. Consequently, the SNR gap between the proposed GLRT detectors and the Conv-ED scales as $\mathcal{O}\big(\sqrt{\frac{MN}{K}}\big)$.
These results highlight the increasing importance of the proposed GLRT detectors in wideband sensing scenarios where $N \gg K$, as well as in high-dimensional spatial regimes where $M \gg 1$.
%
%

\bibliographystyle{IEEEtran}
\bibliography{references}

@ARTICLE{GLRT_paper,
  author={Rasteh, Ali and Rangan, Sundeep},
  journal={IEEE Open J. Commun. Soc.}, 
  title={Computationally Efficient Signal Detection With Unknown Bandwidths}, 
  year={2026},
  volume={7},
  number={},
  pages={997-1018},
  doi={10.1109/OJCOMS.2026.3656868}}

@ARTICLE{Survey_1,
  author={Ali, Abdelmohsen and Hamouda, Walaa},
  journal={IEEE Commun. Surveys Tuts.}, 
  title={Advances on Spectrum Sensing for Cognitive Radio Networks: Theory and Applications}, 
  year={2017},
  volume={19},
  number={2},
  pages={1277-1304},
  doi={10.1109/COMST.2016.2631080}}

@ARTICLE{Survey_2,
  author={Yucek, Tevfik and Arslan, Huseyin},
  journal={IEEE Commun. Surveys Tuts.}, 
  title={A survey of spectrum sensing algorithms for cognitive radio applications}, 
  year={2009},
  volume={11},
  number={1},
  pages={116-130},
  doi={10.1109/SURV.2009.090109}}

@ARTICLE{Survey_3,
  author={Axell, Erik and Leus, Geert and Larsson, Erik G. and Poor, H. Vincent},
  journal={IEEE Signal Process. Mag.}, 
  title={Spectrum Sensing for Cognitive Radio : State-of-the-Art and Recent Advances}, 
  year={2012},
  volume={29},
  number={3},
  pages={101-116},
  doi={10.1109/MSP.2012.2183771}}

@ARTICLE{ED_1,
  author={Margoosian, Argin and Abouei, Jamshid and Plataniotis, Konstantinos N.},
  journal={IEEE Trans. Signal Process.}, 
  title={An Accurate Kernelized Energy Detection in {G}aussian and non-{G}aussian/Impulsive Noises}, 
  year={2015},
  volume={63},
  number={21},
  pages={5621-5636},
  doi={10.1109/TSP.2015.2457400}}

@ARTICLE{ED_5,
  author={Sobron, Iker and Diniz, Paulo S. R. and Martins, Wallace A. and Velez, Manuel},
  journal={IEEE Trans. Commun.}, 
  title={Energy Detection Technique for Adaptive Spectrum Sensing}, 
  year={2015},
  volume={63},
  number={3},
  pages={617-627},
  doi={10.1109/TCOMM.2015.2394436}}

@ARTICLE{ED_14,
  author={Axell, Erik and Larsson, Erik G.},
  journal={IEEE J. Sel. Areas Commun.}, 
  title={Optimal and Sub-Optimal Spectrum Sensing of {OFDM} Signals in Known and Unknown Noise Variance}, 
  year={2011},
  volume={29},
  number={2},
  pages={290-304},
  doi={10.1109/JSAC.2011.110203}}

@ARTICLE{cyclo_2,
  author={Huang, Guangjie and Tugnait, Jitendra K.},
  journal={IEEE Trans. Signal Process.}, 
  title={On Cyclostationarity Based Spectrum Sensing Under Uncertain {G}aussian Noise}, 
  year={2013},
  volume={61},
  number={8},
  pages={2042-2054},
  doi={10.1109/TSP.2013.2246158}}

@ARTICLE{cyclo_3,
  author={Rebeiz, Eric and Urriza, Paulo and Cabric, Danijela},
  journal={IEEE Trans. Signal Process.}, 
  title={Optimizing Wideband Cyclostationary Spectrum Sensing Under Receiver Impairments}, 
  year={2013},
  volume={61},
  number={15},
  pages={3931-3943},
  doi={10.1109/TSP.2013.2262680}}

@ARTICLE{cov_2,
  author={Bianchi, P. and Debbah, M. and Maida, M. and Najim, J.},
  journal={IEEE Trans. Inf. Theory}, 
  title={Performance of Statistical Tests for Single-Source Detection Using Random Matrix Theory}, 
  year={2011},
  volume={57},
  number={4},
  pages={2400-2419},
  doi={10.1109/TIT.2011.2111710}}

@ARTICLE{cov_3,
  author={Wang, Pu and Fang, Jun and Han, Ning and Li, Hongbin},
  journal={IEEE Trans. Veh. Technol.}, 
  title={Multiantenna-Assisted Spectrum Sensing for Cognitive Radio}, 
  year={2010},
  volume={59},
  number={4},
  pages={1791-1800},
  doi={10.1109/TVT.2009.2037912}}

@ARTICLE{cov_4,
  author={Zeng, Y. and Liang, Y.-C.},
  journal={IEEE Trans. Commun.}, 
  title={Eigenvalue-based spectrum sensing algorithms for cognitive radio}, 
  year={2009},
  volume={57},
  number={6},
  pages={1784-1793},
  doi={10.1109/TCOMM.2009.06.070402}}

@ARTICLE{cov_5,
  author={Zeng, Yonghong and Liang, Ying-Chang},
  journal={IEEE Trans. Veh. Technol.}, 
  title={Spectrum-Sensing Algorithms for Cognitive Radio Based on Statistical Covariances}, 
  year={2009},
  volume={58},
  number={4},
  pages={1804-1815},
  doi={10.1109/TVT.2008.2005267}}

@ARTICLE{cov_6,
  author={Kortun, Ayse and Ratnarajah, Tharmalingam and Sellathurai, Mathini and Zhong, Caijun and Papadias, Constantinos B.},
  journal={IEEE J. Sel. Topics Signal Process.}, 
  title={On the Performance of Eigenvalue-Based Cooperative Spectrum Sensing for Cognitive Radio}, 
  year={2011},
  volume={5},
  number={1},
  pages={49-55},
  doi={10.1109/JSTSP.2010.2066957}}

@ARTICLE{subNyquist_1,
  author={Yen, Chia-Pang and Tsai, Yingming and Wang, Xiaodong},
  journal={IEEE Trans. Signal Process.}, 
  title={Wideband Spectrum Sensing Based on Sub-{N}yquist Sampling}, 
  year={2013},
  volume={61},
  number={12},
  pages={3028-3040},
  doi={10.1109/TSP.2013.2251342}}

@ARTICLE{subNyquist_3,
  author={Cohen, Deborah and Eldar, Yonina C.},
  journal={IEEE Trans. Signal Process.}, 
  title={Sub-{N}yquist Sampling for Power Spectrum Sensing in Cognitive Radios: A Unified Approach}, 
  year={2014},
  volume={62},
  number={15},
  pages={3897-3910},
  doi={10.1109/TSP.2014.2331613}}

@ARTICLE{subNyquist_4,
  author={Qin, Zhijin and Gao, Yue and Plumbley, Mark D. and Parini, Clive G.},
  journal={IEEE Trans. Signal Process.}, 
  title={Wideband Spectrum Sensing on Real-Time Signals at Sub-{N}yquist Sampling Rates in Single and Cooperative Multiple Nodes}, 
  year={2016},
  volume={64},
  number={12},
  pages={3106-3117},
  doi={10.1109/TSP.2015.2512562}}

@ARTICLE{WB_1,
  author={Quan, Zhi and Cui, Shuguang and Sayed, Ali H. and Poor, H. Vincent},
  journal={IEEE Trans. Signal Process.}, 
  title={Optimal Multiband Joint Detection for Spectrum Sensing in Cognitive Radio Networks}, 
  year={2009},
  volume={57},
  number={3},
  pages={1128-1140},
  doi={10.1109/TSP.2008.2008540}}

@ARTICLE{WB_3,
  author={Ali, Abdelmohsen and Hamouda, Walaa},
  journal={IEEE Trans. Commun.}, 
  title={Generalized {FFT}-Based One-Bit Quantization System for Wideband Spectrum Sensing}, 
  year={2020},
  volume={68},
  number={1},
  pages={82-92},
  doi={10.1109/TCOMM.2019.2946814}}

@ARTICLE{WB_4,
  author={Lu, Lu and Wu, Hsiao-Chun and Iyengar, S. S.},
  journal={IEEE J. Sel. Areas Commun.}, 
  title={A Novel Robust Detection Algorithm for Spectrum Sensing}, 
  year={2011},
  volume={29},
  number={2},
  pages={305-315},
  doi={10.1109/JSAC.2011.110204}}

@ARTICLE{wavelet_5,
  author={Zhen, Pan and Zhang, Bangning and Chen, Zhibo and Guo, Daoxing and Ma, Wenfeng},
  journal={IEEE Wireless Commun. Lett.}, 
  title={Spectrum Sensing Method Based on Wavelet Transform and Residual Network}, 
  year={2022},
  volume={11},
  number={12},
  pages={2517-2521},
  doi={10.1109/LWC.2022.3207296}}

@ARTICLE{wavelet_7,
  author={Moulin, P.},
  journal={IEEE Trans. Signal Process.}, 
  title={Wavelet thresholding techniques for power spectrum estimation}, 
  year={1994},
  volume={42},
  number={11},
  pages={3126-3136},
  doi={10.1109/78.330372}}

@ARTICLE{sweep_1,
  author={Flak, Przemysław},
  journal={IEEE Access}, 
  title={Hardware-Accelerated Real-Time Spectrum Analyzer With a Broadband Fast Sweep Feature Based on the Cost-Effective {SDR} Platform}, 
  year={2022},
  volume={10},
  number={},
  pages={110934-110946},
  doi={10.1109/ACCESS.2022.3215800}}

@ARTICLE{sweep_2,
  author={Torlak, Murat and Namgoong, Won},
  journal={IEEE Access}, 
  title={Spectral Detection of Frequency-Sparse Signals: Compressed Sensing vs. Sweeping Spectrum Scanning}, 
  year={2021},
  volume={9},
  number={},
  pages={30060-30070},
  doi={10.1109/ACCESS.2021.3053957}}

@ARTICLE{filter_bank_1,
  author={Dikmese, Sener and Sofotasios, Paschalis C. and Renfors, Markku and Valkama, Mikko},
  journal={IEEE Trans. Signal Process.}, 
  title={Subband Energy Based Reduced Complexity Spectrum Sensing Under Noise Uncertainty and Frequency-Selective Spectral Characteristics}, 
  year={2016},
  volume={64},
  number={1},
  pages={131-145},
  doi={10.1109/TSP.2015.2480048}}

@ARTICLE{filter_bank_2,
  author={Farhang-Boroujeny, Behrouz},
  journal={IEEE Trans. Signal Process.}, 
  title={Filter Bank Spectrum Sensing for Cognitive Radios}, 
  year={2008},
  volume={56},
  number={5},
  pages={1801-1811},
  doi={10.1109/TSP.2007.911490}}

@ARTICLE{dyadic_1,
  author={Arias-Castro, E. and Donoho, D.L. and Xiaoming Huo},
  journal={IEEE Trans. Inf. Theory}, 
  title={Near-optimal detection of geometric objects by fast multiscale methods}, 
  year={2005},
  volume={51},
  number={7},
  pages={2402-2425},
  doi={10.1109/TIT.2005.850056}}

@INPROCEEDINGS{freq_hop,
  author={Hasan, Md. Zoheb and Couto, David J. and Abdel-Malek, Mai A. and Reed, Jeffrey H.},
  booktitle={2023 IEEE 34th Annu. Int. Symp. Pers. Indoor Mobile Radio Commun. (PIMRC)}, 
  title={Frequency Hopping Signal Detection in Low Signal-to-Noise Ratio Regimes}, 
  year={2023},
  volume={},
  number={},
  pages={1-7},
  doi={10.1109/PIMRC56721.2023.10293834}}

@ARTICLE{ML_1,
  author={Sarikhani, Rahil and Keynia, Farshid},
  journal={IEEE Commun. Lett.}, 
  title={Cooperative Spectrum Sensing Meets Machine Learning: Deep Reinforcement Learning Approach}, 
  year={2020},
  volume={24},
  number={7},
  pages={1459-1462},
  doi={10.1109/LCOMM.2020.2984430}}

@ARTICLE{ML_2,
  author={Tian, Jinfeng and Cheng, Peng and Chen, Zhuo and Li, Mingqi and Hu, Honglin and Li, Yonghui and Vucetic, Branka},
  journal={IEEE Trans. Veh. Technol.}, 
  title={A Machine Learning-Enabled Spectrum Sensing Method for {OFDM} Systems}, 
  year={2019},
  volume={68},
  number={11},
  pages={11374-11378},
  doi={10.1109/TVT.2019.2943997}}

@ARTICLE{ML_3,
  author={Shi, Zhenjiang and Gao, Wei and Zhang, Shangwei and Liu, Jiajia and Kato, Nei},
  journal={IEEE Trans. Wireless Commun.}, 
  title={Machine Learning-Enabled Cooperative Spectrum Sensing for Non-Orthogonal Multiple Access}, 
  year={2020},
  volume={19},
  number={9},
  pages={5692-5702},
  doi={10.1109/TWC.2020.2995594}}

@ARTICLE{ML_4,
  author={Abusubaih, Murad A. and Khamayseh, Sundous},
  journal={IEEE Access}, 
  title={Performance of Machine Learning-Based Techniques for Spectrum Sensing in Mobile Cognitive Radio Networks}, 
  year={2022},
  volume={10},
  number={},
  pages={1410-1418},
  doi={10.1109/ACCESS.2021.3138888}}

@book{Szego,
  title={Toeplitz forms and their applications},
  author={Grenander, Ulf and Szeg{\"o}, Gabor},
  year={1958},
  publisher={Berkeley, CA, USA: Univ. Calif. Press}
}

@ARTICLE{toeplitz,
  author={Mei, Ruru and Wang, Zhugang},
  journal={Found. Trends Commun. Inf. Theory}, 
  title={Toeplitz and circulant matrices: A review}, 
  year={2005},
  volume={2},
  number={3},
  pages={155–239}}

@ARTICLE{TAES_ED_1,
  author={Lipski, Michael V. and Kompella, Sastry and Narayanan, Ram M.},
  journal={IEEE Trans. Aerosp. Electron. Syst.}, 
  title={Practical Implementation of Adaptive Threshold Energy Detection using Software Defined Radio}, 
  year={2021},
  volume={57},
  number={2},
  pages={1227-1241},
  doi={10.1109/TAES.2020.3040059}}

@ARTICLE{TAES_ED_2,
  author={Mattingly, Rylee G. and Metcalf, Justin G.},
  journal={IEEE Trans. Aerosp. Electron. Syst.}, 
  title={Fast Adaptive Spectrum Sensing Using Hardware Optimized, Cell Averaging Estimation for Cognitive Radio and Radar Applications}, 
  year={2023},
  volume={59},
  number={5},
  pages={5084-5096},
  doi={10.1109/TAES.2023.3252500}}

@ARTICLE{TAES_FFT,
  author={Cheng, CHI-HAO and Liou, Lihyeh L. and Tsui, James B. and Lin, David M.},
  journal={IEEE Trans. Aerosp. Electron. Syst.}, 
  title={Chirp signal detection using {FFT} peak frequency difference [Correspondence]}, 
  year={2016},
  volume={52},
  number={3},
  pages={1449-1453},
  doi={10.1109/TAES.2016.140201}}

@ARTICLE{CUSUM_1,
  author={Tartakovsky, A.G. and Rozovskii, B.L. and Blazek, R.B. and Hongjoong Kim},
  journal={IEEE Trans. Signal Process.}, 
  title={A novel approach to detection of intrusions in computer networks via adaptive sequential and batch-sequential change-point detection methods}, 
  year={2006},
  volume={54},
  number={9},
  pages={3372-3382},
  doi={10.1109/TSP.2006.879308}}

@book{CUSUM_2,
  title={Sequential analysis: Hypothesis testing and changepoint detection},
  author={Tartakovsky, Alexander and Nikiforov, Igor and Basseville, Michele},
  year={2014},
  publisher={CRC press}
}

\end{document}